# Duality of wave modulation and nanotwinning in Ni–Mn–Ga martensite via long-period commensurate states

P. Veřtát[1], M. Zelený[2], A. Sozinov[3], M. Klicpera[4], O. Fabelo[5], R. Chulist[6], M. Vinogradova[3], P. Sedlák[7], H. Seiner[7], O. Heczko[1], and L. Straka[1*]

[1] *FZU – Institute of Physics of the Czech Academy of Sciences, Prague, Czechia*

[2] *Faculty of Mechanical Engineering, Institute of Materials Science and Engineering, Brno University of Technology, Brno, Czechia*

[3] *Material Physics Laboratory, Lappeenranta-Lahti University of Technology LUT, Lappeenranta, Finland*

[4] *Department of Condensed Matter Physics, Faculty of Mathematics and Physics, Charles University, Prague, Czechia*

[5] *Institut Laue-Langevin, Grenoble, France*

[6] *Faculty of Metals Engineering and Industrial Computer Science, AGH University of Krakow, Krakow, Poland*

[7] *Institute of Thermomechanics of the Czech Academy of Sciences, Prague, Czechia*

*Corresponding author: Ladislav Straka, FZU – Institute of Physics of the Czech Academy of Sciences, Na Slovance 1999/2, 182 00 Prague 8, Czechia. Email: ladislav.straka@fzu.cz, phone: +420 266 052 992

**Abstract**

Structural modulation is a key ingredient behind the extraordinary (magneto)elastic response of Ni–Mn–Ga martensite, yet its link to fine microstructural features and twin-boundary supermobility remains unresolved. Here we analyse martensitic single crystals of $Ni_{50.0}Mn_{27.7}Ga_{22.3}$ and $Ni_{50.0}Mn_{28.1}Ga_{21.9}$. Neutron and X-ray diffraction reveal an anharmonic five-layer structural modulation—evidenced by high-order satellite reflections—that evolves from commensurate ($q$ = 2/5) to incommensurate (2/5 < $q$ < 5/12) upon cooling. Interpreting the refined modulation displacements as a basal-plane stacking sequence links the wave description to the microstructural evolution on cooling. In this view, evolving incommensurability produces periodic nanodomains interpreted as emerging *a/b*-nanotwins with a characteristic size of ≈20 nm at ≈290 K. With further cooling, the modulation can lock into long-period commensurate (LP-C) states, such as 34O ($q$ = 7/17), 24O ($q$ = 5/12), and 14O ($q$ = 3/7), whose orthorhombic unit cells can be viewed as *a/b*-nanotwins. Ab initio calculations show that LP-C structures are energetically competitive with the initial commensurate state, supporting a shallow martensitic energy landscape. We propose a physical picture in which the martensitic transformation selects a commensurate state with $q$ = 2/5 in the Mn-rich compositions studied here, while subsequent cooling drives relaxation within the martensitic landscape toward LP-C states, particularly 24O in the present alloys. The resulting structure is neither purely wave-like nor purely nanotwinned; rather, it reflects coupling between a coherent modulation wave and local accommodation via NM-like tetragonal distortions, nanotwinning, and LP-C lock-ins, providing a structural basis for the wave–nanotwin duality in Ni–Mn–Ga martensite.

## 1 Introduction

The modulated crystal structures observed in Ni–Mn–Ga alloys exhibiting magnetic shape memory behavior are fundamental to their distinct properties, particularly their high twin-boundary mobility [1–6]. This mobility enables functionalities such as magnetic field-induced strain and related magnetoelastic effects, motivating applications in actuation, sensing, and energy harvesting [7–9]. A striking manifestation is twin-boundary *supermobility,* i.e., unusually large mobility across a broad temperature window from the martensitic transformation to liquid-helium temperature [5,6].

Operationally, the supermobility can be understood as twin-boundary motion at twinning stresses far below the magnetic driving stress available from magnetocrystalline anisotropy, typically below about one tenth of the ≈3 MPa magnetic driving stress, i.e., $\sigma_{TW} < 0.3$ MPa [10]. For comparison, martensite reorientation/detwinning stresses in conventional NiTi shape-memory alloys are typically two to three orders of magnitude higher, in the tens to hundreds of MPa range [4]. Twin-boundary supermobility in Ni–Mn–Ga is associated with an enormous shear elastic instability in the lattice (elastic anisotropy ratio $A^*$ up to 255 [11]) and has been linked to the presence of structural modulation and local, stress-induced defects in the modulation stacking sequence [6,12–14].

The dynamical origin of modulation is commonly linked to anomalous softening of the [ξξ0] $TA_2$ phonon branch in austenite and to its condensation during the transformation toward modulated martensite [15,16]. However, the static structure—how the modulation is accommodated in the martensitic lattice—and its evolution on cooling remain less well understood. This gap is particularly relevant for supermobility:

the way in which the modulation controls twin-boundary (super)mobility across temperature and composition remains unresolved, motivating a closer look at the modulation character and evolution.

The most practically relevant and most extensively studied structural case is the five-layer modulated martensite (10M when chemical ordering is considered; 5M otherwise) [1,17–19]. It combines exceptionally high twin-boundary mobility [2,20] with large field-driven actuation strains and demonstrated device-level use [8,9]. In this phase, the modulation is commonly described as commensurate (C) at $q = 2/5$ and incommensurate (IC) for $q \neq 2/5$ (typically $q \geq 2/5$ in Ni–Mn–Ga), where $q$ denotes the scalar component of the modulation vector along the $[110]^*$ direction (using the *cubic* $L2_1$-frame coordinates) [21–25]. A concise overview of reported five-layer structural forms and notation conventions is provided in Appendix A. Figure 1 schematically summarizes representative modulation cases and the commonly used nanotwinning description based on the adaptive martensite concept [12,26,14].

A long-standing issue is the apparent duality in structural descriptions of the modulation [14]. Diffraction data commonly support a continuous displacement wave, Fig. 1a–d, inferred from satellites indexed by a single modulation vector [22,25], whereas high-resolution atomic imaging and *ab initio* calculations often support a lattice built from nanotwinned blocks [14,27], Fig. 1e. In addition, composition- and temperature-dependent changes in the modulation vector component $q$ and the average symmetry have been reported [23–25], but the mechanisms by which the changes in $q$ drive symmetry transformations and the accompanying reorganization of the twinned microstructure remain unclear.

Another key aspect concerns commensurate lock-in and long-period commensurate (LP-C) states, which are common features of many modulated systems (see Appendix A) [28,29]; it is therefore plausible that they also occur in Ni–Mn–Ga. However, LP-C states in this system have not been systematically examined. We therefore study the evolution of $q$ beyond 2/5 with temperature and composition, focusing on identifying the LP-C states and assessing their role as potential intermediate states along the modulation pathway.

Clarifying these links between modulation, nanotwinning, LP-C states, and symmetry provides the context for the detailed experimental and theoretical investigations presented here. In this work, we aim to:

- build on the previously established anharmonic modulation function [30] to analyse the structural evolution of 10M martensite,
- relate the evolution of the modulation parameter $q$ to the reorganization of the twinned microstructure, and
- identify and physically interpret LP-C states that arise during the evolution of the modulation.

To address these aims, we first re-examine the anharmonic modulation in two exceptional compositions that retain a five-layer modulation at low temperatures. Then we show how the evolving lattice incommensurability leads to the formation of nanodomains, changes in lattice symmetry, and the emergence of LP-C states. Ab initio (DFT+U) calculations indicate that these LP-C states can be energetically competitive with the short-period modulation, while literature data allow us to identify distinct LP-C structures in commonly studied compositions. As the last step, we discuss a physical picture of $q$ evolution and LP-C lock-in, emphasizing the complementary roles of wave modulation and nanotwinning in the evolving martensitic structure.

## 2 Experimental

### *2.1 Sample preparation and microstructure simplification*

Single crystals were grown at AdaptaMat Ltd. by a directional solidification from high-purity raw components. The compositions were determined using X-ray fluorescence spectroscopy as $Ni_{50.0}Mn_{27.7}Ga_{22.3}$ and $Ni_{50.0}Mn_{28.1}Ga_{21.9}$ (at. %) for alloys 1 and 2, respectively. Samples, extracted from heat-treated ingots, were cut into rectangular parallelepipeds (2.5 × 1 × 10 mm³ and 2.5 × 3 × 10 mm³) with faces oriented approximately parallel to the {100} planes of the parent cubic austenite. They were mechanically ground and subsequently electropolished as the final preparation step. The transformation temperatures were determined using DC magnetic susceptibility measurements: $M_s \approx M_f \approx 309$ K, $A_s \approx A_f \approx 315$ K for alloy 1 and $M_s \approx M_f \approx 297$ K, $A_s \approx A_f \approx 301$ K for alloy 2.

For the present study, the twin microstructure was deliberately simplified (partially detwinned) by careful sample selection and by applying a small mechanical pre-compression. Samples that contained a single dominant modulation domain were chosen for the study (see Ref. [31] for details of modulation-domain reorientation). Then, prior to neutron diffraction (ND) and X-ray diffraction (XRD) experiments, the samples were compressed under a small stress (<3 MPa) along the long geometrical axis. This procedure stabilizes a state with the short *c*-axis aligned with the compression direction [32], while it preserves a single dominant modulation direction. As a result, during diffraction measurements the samples exhibited a simplified twin configuration: they were virtually free of *a/c*-twins (no switching between *a*- and *c*-axes) and showed substantially reduced modulation twinning, with one modulation direction being strongly dominant. Thus, only *a/b*-twins remained, together with a single modulation direction, allowing us to focus exclusively on the relationship between the evolution of the modulation period and *a/b*-twinning.

### *2.2 Diffraction experiments and patterns calculation*

The neutron diffraction (ND) experiments were conducted at ILL Grenoble using two four-circle neutron diffractometers equipped with He cryostats. The $Ni_{50}Mn_{27.7}Ga_{22.3}$ sample was measured using the D10 instrument employing a wavelength of $\lambda$ = 0.2360 nm [33], and the $Ni_{50}Mn_{28.1}Ga_{21.9}$ sample was measured using the D9 instrument with $\lambda$ = 0.0838 nm [34]. Two-dimensional (2D) area detectors were utilized to accommodate small sample orientation changes during cooling and heating. In data analysis, the contributions from minor martensite variants and background noise were minimized by restricting the 2D integration area. A detailed description of the data processing methodology is available in the Supplementary Material of a related publication [30].

Unless otherwise stated, we use the *cubic ($L2_1$-frame)* lattice coordinates throughout this article. In selected cases, we also employ *diagonal* coordinates, i.e., the frame of the smallest face-centered tetragonal unit cell (often denoted $L1_0$). We use both coordinate systems because the cubic frame is physically most convenient for comparison with the austenite lattice and for discussing functional properties, whereas the diagonal frame is better suited for describing and modeling the modulation, which is aligned with a principal axis in this setting. See inset in Fig. 2c and Appendix A1 for further details.

The X-ray diffraction (XRD) measurements were performed using a Bruker D8 Discover diffractometer with a rotating Cu anode ($\lambda$ = 0.1542 nm) equipped with Anton Paar DCS 350 cooling stage and a point detector. Reciprocal-space mapping was conducted to confirm that a single modulation direction was dominant, and scans included reflections such as (400) and (620) together with their associated modulation satellites.

Following precise sample orientation alignment in the diffractometer, one-dimensional scans along $[110]^*$ direction (q-scans) were performed. For ND, the q-scans included the (2-20) and (400) reflections and the enclosed satellites, and for XRD, the scans included the (400) and (620) reflections and the respective satellites. Modulation was evaluated using both reciprocal space maps (XRD only) and q-scans (ND and XRD) at various temperatures.

The indexing of reflections followed previous studies by Fukuda et al. [35], Mariager et al. [36], and Righi et al. [24], using the established wave modulation approach. The magnitude of the modulation vector component $q$ was determined based on the relative spacings of principal and satellite reflections. This approach ensures $q$ is unaffected by thermal expansion because the modulation is confined to the $[110]^*$ direction. To account for minor instrumental factors like small sample displacements or orientation drifts during temperature changes, the satellite spacings were averaged to determine the value of $q$.

The DISCUS simulation package [37] was employed to calculate the diffraction patterns corresponding to the reference modulation function (Eq. 2). Simulations were performed for $Ni_{50}Mn_{25}Ga_{25}$ by averaging 20 crystal models, each comprising 2 × 1 × 470 base unit cells defined in diagonal coordinates, with lattice parameters 4.22 × 5.65 × 4.21 Å. Calculations with increased Mn content ($Ni_{50}Mn_{28}Ga_{22}$) showed no significant effect on the calculated diffraction patterns along the modulation direction.

### *2.3 Ab initio calculations*

The ab initio calculations based on DFT+U were performed using the Vienna Ab initio Simulation Package (VASP) [38,39] with projector-augmented wave (PAW) potentials [40,41]. The electronic orbitals were expanded using plane waves with a maximum kinetic energy of 800 eV, explicitly including the $3p^6 3d^9 4s^1$ electrons for Ni, $3p^6 3d^6 4s^1$ for Mn, and $3d^{10} 4s^2 4p^1$ for Ga as valence states. The Brillouin zone was sampled using a Γ-point-centered mesh with a minimum k-point spacing of 0.08 $Å^{-1}$. The Methfessel-Paxton smearing method [42] (smearing width parameter: 0.07 eV) was used for initial relaxations, followed by recalculations of total energies using the modified tetrahedron method [43] to ensure precision. The gradient-corrected exchange-correlation functional (GGA) proposed by Perdew, Burke, and Ernzerhof [44] was employed, including non-spherical contributions inside the PAW spheres. The rotationally invariant DFT+U method proposed by Dudarev [45] was applied to address electron localization.

The calculations were performed for the stoichiometric $Ni_{50}Mn_{25}Ga_{25}$ composition to avoid the complex magnetic interactions and potential spin frustrations caused by randomly distributed Mn-excess atoms. Although some quantitative differences between the stoichiometric model and the Mn-rich experimental compositions are expected, we assume that the qualitative trends in the relative stability of the modulated structures remain representative.

Input computational cells with orthorhombic lattices for modulated martensites were generated using the anharmonic modulation function described in Eq. (2). Full relaxation of atomic positions and structural parameters was carried out using the quasi-Newton algorithm, with convergence criteria set to an energy change $< 0.01$ meV/atom. To minimize errors due to cell size differences, energies of modulated structures were always calculated relative to those of non-modulated martensite in cells with identical numbers of atoms. Convergence tests confirmed that this setup provided total energy estimates with an accuracy better than 0.1 meV/atom.

Previous studies showed that DFT results without electron-localization correction (U) deviated significantly from experiments. Most notably, the calculated 10M structure exhibited overestimated monoclinicity and significantly different *a* and *b* axes (≈5% difference in uncorrected calculation in contrast to ≈0.5% in the experiment) [46]. Furthermore, a 4O structure emerged as the most stable despite never being experimentally observed in Ni–Mn–Ga [47]. This highlights a known limitation of standard DFT (U=0) for Ni–Mn–Ga.

A range of U values from 0.50 eV to 5.97 eV for Mn atoms has been proposed. However, large U suppresses the martensitic transformation as the cubic parent phase of austenite becomes the most stable structure [48]. The best agreement between calculated and experimental lattice parameters was reached for U = 1.8 eV [46], whereas Bhattacharya et al. [49] found that U = 0.5 eV provides the best agreement between calculated and experimentally measured hard X-ray photoelectron spectra of 14M martensite. Therefore, we calculate the energy landscape for U from 0 to 1.8 eV, without attempting to identify a single optimal value, which is beyond the scope of this work. Our goal is to assess the energetic competitiveness of LP-C structures across a range of U values, bearing in mind that the present DFT+U results are 0 K total energies of idealized commensurate approximants and therefore cannot, by themselves, uniquely determine the experimentally realized sequence of modulated states at finite temperature.

## 3 Results and Discussion

### *3.1 Anharmonic incommensurate modulation − diffraction and reference function*

This section presents our neutron and X-ray diffraction experiments on single crystals and, by comparison with calculated diffraction patterns, demonstrates that the modulation is well described by the anharmonic modulation function determined in Ref. [30], hereafter referred to as the *reference modulation function*. We also show that the continuous modulation wave can be qualitatively represented as a stacking sequence of basal planes, a viewpoint that will later be used to describe modulation-induced microstructural features.

Figure 2a presents a schematic phase diagram of the $Ni_{50}Mn_{25+x}Ga_{25-x}$ (at. %) system, based on Ref. [50]. The measurement paths for the studied compositions are marked, clarifying each alloy's compositional position and the projected phase evolution. The selected compositions $x = 2.7$ ($Ni_{50.0}Mn_{27.7}Ga_{22.3}$, alloy 1) and $x = 3.1$ ($Ni_{50.0}Mn_{28.1}Ga_{21.9}$, alloy 2) show no intermartensite transformations, allowing a focused investigation of the five-layered martensite down to 2 K.

Figure 2b shows examples of neutron diffraction (ND) q-scans for alloy 2 at various temperatures. It also illustrates that the spacing of satellite reflections $s_{\pm1}$, $s_{\pm2}$, ... directly determines the modulation vector component $q$. For each temperature, $q$ was determined from the average spacing of the fundamental and all resolved satellite reflections (including high-order satellites for the IC structure) detected in the individual q-scans. Figures 2c and 2d present the determined temperature-dependent evolution of $q$ in alloys 1 and 2, respectively. Both alloys are nearly commensurate (C) at room temperature near the martensite transformation, with $q \approx 2/5$, which converges to $q \approx 0.416$ at low temperatures.

A key observation is the presence of numerous high-order satellites in the IC structure (Fig. 2b), which are even more apparent in the XRD patterns (Fig. 3) thanks to lower instrumental broadening. Such a rich satellite landscape indicates that the modulation is not a pure sine function. In diffraction from modulated crystals, higher-order satellites reflect higher Fourier components of the modulation, and thus provide direct information on its spectral content [51–53,30]. In practice, the modulation is therefore often parameterized as a Fourier series, e.g.,

$$dx = \sum_{n=1}^{M} A_n \sin\left(2\pi n q z + \phi_n\right) \qquad (1)$$

where $dx$ is the displacement at coordinate $z$, $A_n$ and $\phi_n$ are the amplitude and phase of the $n$-th harmonic, and $M$ is the highest harmonic retained in the refinement.

The refinements typically include harmonics up to the highest satellite order that is reliably observed (within the limits set by data quality and parameter correlations) [51,52]. In the present work, we do not refine an unconstrained set $\{A_n, \phi_n\}$; instead, we adopt the previously established reference anharmonic modulation function [30]:

$$dx = A_1 \cdot \sin\left(2\pi z/p\right) + 0.2 \cdot A_1 \cdot \sin\left(5 \cdot 2\pi z/p\right) + 0.04 \cdot A_1 \cdot \sin\left(8 \cdot 2\pi z/p\right), \text{ with } p = 2/q \qquad (2)$$

Specifically, we retain only the $n$ = 1, 5, and 8 harmonics with fixed amplitude ratios (and zero phases), so that the refinement reduces to determining only $A_1$ and $p$ (via $p = 2/q$) for each diffraction pattern. The energetic origin of the zero phase of the fundamental harmonic is discussed in Ref. [54]. The displacement $dx$ refers to the shift of the whole (110) plane along [1-10], i.e. we use a rigid-plane approximation. This is similar to the previously used model [55], which also treats basal planes as collectively shifted units, but contrasts with some previous works where the displacement $dx$ of each chemical element was treated independently, e.g. [24,56,57]. The displacement is transverse to the modulation direction, which defines the *z* axis (plane index along the modulation wavevector, see also inset in Fig. 2c for coordinate definition and illustrative overview). The substantial fifth-harmonic term (≈20% of the fundamental amplitude) explains the strong fifth-order satellites (Figure 3d), whereas the inclusion of a smaller eighth-harmonic term is supported by the non-negligible eighth-order satellite intensity and improves the agreement between calculated and experimental diffraction patterns.

Detecting satellites up to the eighth order far exceeds typical experimental observations in modulated crystals, where even third-order satellites are rare, and measurable intensity from satellites beyond the fourth order is exceptionally uncommon [52,58]. Singh et al. [59] noted that a conventional laboratory-source powder XRD setup cannot detect the second and third-order satellites of the martensite phase of $Ni_2MnGa$. Moreover, including higher-order harmonics in the refinement process leads to highly correlated parameters, which can cause the refinement to fail [52]. Consequently, conventional methods may not reveal the strongly anharmonic character of the modulation. Therefore, high-resolution and high-intensity techniques—such as synchrotron-based and single-crystal diffraction—are essential for accurate determination of the harmonic components of the modulation function.

The evaluation of neutron diffraction patterns for the single crystal of alloy 1 and their comparison with model calculations have been presented in previous work [30]. Here we show a complementary result: Figure 3 compares selected experimental X-ray diffraction (XRD) patterns for single crystals of alloy 2 with diffraction patterns calculated using the reference modulation function according to Eq. (2) ($A_1$ and $q$ refined to the experimental patterns). The fit is excellent and similarly strong fits were obtained for both XRD and ND at all measured temperatures for both alloys, as well as for other Ni–Mn–Ga and Ni–Mn–Ga–Fe alloys not discussed here. These consistently high-quality fits support the validity of the reference modulation function given by Eq. (2) and provide a robust foundation for further analysis of the resulting lattice distortions.

Based on these refinements, some variations in the modulation function amplitude ($A_1$) were observed among different alloys and experiments. They can be attributed to various experimental factors such as sample history, composition, and intensity contributions from non-dominant (residual) modulation variants. At room temperature, $A_1$ was between 5–7% (≈30 pm) of the $d_{220}$ plane spacing, rising to 8–10% (≈40 pm) at 10 K. Previous studies have reported comparable modulation amplitudes of 6–10% [22,60].

While modulation is often described as a displacement wave, an alternative and equally insightful approach is to interpret it as a stacking sequence of basal planes or even larger structural blocks [61,62], such as adaptive nanotwins (see Figure 1e). We suggest that the continuous modulation in 10M martensite can also be well represented as a $(2\overline{3})_2$ stacking sequence of (110) basal planes, even for IC modulation, as will be shown below. The $(2\overline{3})_2$ denotes a sequence in which the $(2\overline{3})$ unit (two positive and three negative stacking shifts) is repeated twice to align with chemical order. This doubled notation is retained for consistency with prior studies, although chemical ordering is not essential here.

In Figure 4, the interpretation of continuous modulation as a stacking sequence is illustrated for commensurate modulation ($q$ = 2/5). Figure 4a juxtaposes the reference modulation function with the lattice periodicity (vertical lines), demonstrating how the resulting basal-plane displacements emerge from the interplay between the modulation function and the discrete lattice. Figure 4b shows the resulting structure with displacements at the same scale as the lattice, while Figure 4c exaggerates the displacements threefold to better reveal the similarity between the continuous modulation and the stacking sequence/nanotwinning description (compare with Figure 1e). These visualizations underscore that the stacking sequence can naturally arise from continuous modulation rather than constituting an independent structural feature.

Taken together, these findings validate the reference anharmonic modulation function, which accurately captures the rich satellite landscape observed in the diffraction patterns across a broad temperature range of approximately 2–300 K in two compositionally distinct alloys. Further, interpreting the modulation as a stacking sequence of basal planes helps to understand how the atomic arrangement adapts to changing $q$. This will be explored in the next section, where we analyze microstructural changes driven by the decreasing modulation period $p$ (increasing $q$).

### *3.2 Nanodomains and emerging a/b-nanotwins*

This section examines how the thermal evolution of $q$ drives changes in both lattice symmetry and microstructure. We show that incommensurate anharmonic modulation induces nanodomain formation and then we use the stacking-sequence perspective to link these nanodomains to emerging *a/b*-nanotwins.

#### *3.2.1 Nanodomain formation from lattice–modulation mismatch*

Nanodomains arise from a mismatch between the periodicity of the modulation and that of the discrete lattice. Figure 5 demonstrates the concept, beginning with a well-known phenomenon from acoustics in Figure 5a. When two signals of slightly different frequencies $f_1$ and $f_2$ are summed, the resulting signal exhibits beats or an *"amplitude envelope"*, i.e., periodic fluctuations in amplitude with frequency $|f_1 - f_2|$. A similar "envelope" effect also arises when a continuous wave is sampled by a discrete series of pulses, as illustrated in Figure 5b. This is directly analogous to our structural case, where the continuous modulation wave is sampled by the discrete sequence of lattice planes.

Figure 5c illustrates the effect for our specific case of five-layer modulation. We compare displacements for commensurate modulation ($q$ = 2/5 = 0.400) and incommensurate modulation ($q$ = 0.412). In the commensurate case, atomic displacements exhibit a perfectly repeating pattern every five lattice planes. In contrast, in the incommensurate case, the mismatch in periodicity causes atomic displacements to drift over longer distances, manifesting as the displacement envelope. The envelope period (expressed in lattice planes) therefore defines a characteristic nanodomain size, whose microstructural interpretation is described in the next section.

A larger-scale depiction of the formed domains is given in Figure 5d, highlighting their scaling as a function of $q$. To determine the scaling law, we consider that the domain size corresponds to the number of lattice planes over which the incommensurate modulation slips by one lattice plane relative to the commensurate reference ($p$ = 2/$q$ = 5). One modulation period introduces a cumulative offset of (5−$p$) planes; hence 1/(5−$p$) periods are required to reach a one-plane slip. Multiplying this number of periods by $p$ (planes per period) gives the domain size in lattice planes. Further multiplying by the spacing of successive planes along the modulation direction ($d_{220}$ in the cubic $L2_1$ frame; see also inset in Fig. 2c) converts it to an absolute length:

$$DS = \left(\frac{p}{5-p}\right)\cdot d_{220} = \left(\frac{2}{5q-2}\right)\cdot d_{220} \qquad (p < 5,\ q > 2/5), \tag{3}$$

The domain size is infinitely large in an ideal, exactly commensurate structure ($q$ = 2/5, $p$ = 5). However, even for a very small incommensurability, such as that observed at room temperature, the domain size already reaches the submicron scale ($DS$ $\approx$ 840 nm for $d_{220} \approx 0.21$ nm, $q \approx 0.4001$). As $q$ increases and the modulation period $p$ decreases, the domains get even smaller. Hence, in most cases, the domains are expected to be nanoscopic; we therefore refer to them as *nanodomains* hereafter.

In summary, incommensurate modulation naturally gives rise to nanodomains whose characteristic size follows Eq. (3) and is governed by the deviation from commensurability.

*3.2.2 Emerging a/b-nanotwins from stacking-sequence interpretation*

Next, we link the found nanodomains with discrete structural features, specifically *a/b*-nanotwins, making it easier to understand what these nanodomains represent and how the evolution of modulation is reflected in the microstructure. This is achieved by interpreting the continuous modulation displacements as a basal (110) plane stacking sequence.

A straightforward example of commensurate modulation is detailed in Figures 4 and 5c, illustrating how continuous modulation maps onto a stacking sequence.
A consistent pattern of two positive ("up") shifts and three negative ("down") shifts emerges, corresponding to the specific infinite $(2\overline{3})_2$ stacking sequence. Vinogradova et al. [55] recently showed that the *average lattice* of commensurate modulated Ni–Mn–Ga-Fe martensites, as reflected by the principal diffraction peaks, can be accurately described using the stacking-plane sequence with a *constant plane shift* (CPS model, where $|dx|$ = a constant shift magnitude between neighboring planes). The fitted plane-shift magnitude varies between martensitic structures, indicating that it depends on the modulation state. The CPS model is in line with earlier similar concepts [61,62] and the nanotwinning model of the structure (Figure 1e), but it was not developed for incommensurate structures.

For the present qualitative interpretation of the incommensurate structure, we retain the stacking-sequence perspective used in the CPS model. However, the plane shifts must vary to reproduce the anharmonic incommensurate modulation according to Eq. (2). Qualitatively, each plane shift remains either positive (+1, up) or negative (–1, down), mostly maintaining the $(2\bar{3})_2$-like sequence (Figures 5c and 5d). The key difference is that local $2\bar{2}$-type faults additionally emerge to accommodate the shorter modulation period. These $2\bar{2}$-type faults are marked as $2|\bar{2}$ interfaces, with "|" indicating the mid-plane of the interface or nanotwin boundary plane as described next.

The stacking-sequence perspective enables us to connect the atomic-scale displacements to larger-scale structural features—specifically, we associate the $2|\bar{2}$ interfaces with *a/b*-nanotwin boundaries. The displacements in the incommensurate structure interpreted as a stacking sequence are given in Figure 5c, d. Periodic reversals occur between $(2\bar{3})_2$ and its inverted $(3\bar{2})_2$ stackings, separated by $2|\bar{2}$ interfaces. Such reversals correspond to the structural model of *a/b*-nanotwins as inverting stacking faults [14,63,64]. The "|" symbol then represents the *a/b*-nanotwin boundary plane, separating the two nanotwin domains. Furthermore, beyond its alignment with the inverting stacking sequence model of *a/b*-nanotwins, the $(2\bar{3})_2$ stacking induces a shear, mirrored in the opposite $(3\bar{2})_2$ stacking. This is consistent with the continuum theory view of a compound twin as two oppositely sheared domains. Therefore, the periodicity mismatch between the lattice and the incommensurate modulation naturally leads to the formation of $(2\bar{3})_2 \leftrightarrow (3\bar{2})_2$ stacking inversions, corresponding to *a/b*-nanotwins. In this picture, the nanodomain size derived by Eq. (3) corresponds to the characteristic *a/b*-nanotwin thickness.

However, this sequence-inversion description alone cannot account for the full complexity of nanodomain formation, nor for the preference of particular twinned states such as the LP-C structures discussed later. Additional driving factors are required to bias the lattice toward energetically favorable twin configurations. These two factors are presumably important: i) the extraordinarily low energy of $2|\bar{2}$ interfaces and ii) the low energy of NM martensite.

The role of the $2|\bar{2}$ interfaces is supported by prior ab initio calculations of stacking-fault energies. In the stacking-sequence argument above, $3|\bar{3}$ twin interfaces do not appear and calculations show they are high-energy, whereas the $2|\bar{2}$ interfaces that occur periodically were calculated to be the most energetically favorable. These interfaces correspond to a $(2\bar{2})$ stacking sequence—that is, the 4O structure [47,64]. The second factor involves local relaxation of the modulated structure toward an NM-like configuration, Fig. 1e. NM martensite is both energetically favorable and often regarded as a fundamental building block of the twinned lattice [12,65,66].

Thus, the stacking-sequence interpretation shows that nanodomains introduced in Section 3.2.1 can be viewed as emerging *a/b*-nanotwins, which may then be energetically stabilized into well-defined *a/b*-nanotwins, primarily via the formation of $2|\bar{2}$ boundaries and local relaxation toward NM-like configurations. In this view, the domain size DS from Eq. (3) should be understood as a geometrically defined mismatch (envelope) length that sets the characteristic *a/b*-nanotwin scale, while real twin/domain sizes are expected to form a distribution broadened by defects and local chemical variations (see Fig. 6 and the SEM discussion below). The proposed energetic framework is conceptual, and additional factors may also contribute to nanodomain formation. Nevertheless, it provides a useful theoretical basis for understanding how periodicity-mismatch-driven stacking inversions may develop into the structures that resemble or fully realize *a/b*-nanotwins. Having established this structural interpretation, we next consider experimental indications of the corresponding *a/b*-twin refinement upon cooling.

### *3.2.3 Experimental indications of a/b-twin refinement*

Qualitative SEM observations and indirect XRD analysis provide complementary support for *a/b*-twin refinement upon cooling and for the evolving nanodomains and *a/b*-nanotwinning hypothesis.

The *a/b*-twins are difficult to observe because of their small twinning shear, resulting in minimal misorientation (<0.3°) and an almost undetectable difference in the *a* and *b* lattice parameters ($a/b < 1.005$). Despite these challenges, several studies have directly imaged *a/b*-twins using scanning/transmission electron microscopy (SEM/TEM). Reported methods include direct TEM imaging [14] and manually indexed electron backscattered diffraction (EBSD) in SEM [67]. In backscattered electron (BSE) imaging in SEM [68,69], electron channeling effects enhance visibility. Optical observations of broad twins have also been demonstrated [70].

Using BSE contrast in SEM, we observed the evolution of *a/b*-twin-related contrast with cooling in the $Ni_{50.0}Mn_{28.2}Ga_{21.8}$ alloy, as shown in Figure 6. At 298 K and $q \approx 2/5$, broad *a/b*-twin bands are discernible (Figure 6a). However, at 253 K and $q > 2/5$, significantly finer contrast appears in the same region (Figure 6b). A broader set of observations shows a continuous refinement of contrast upon cooling and a reverse effect, coarsening of contrast upon heating, aligned with the hysteresis of the $q(T)$ dependence in $Ni_{50.0}Mn_{28.2}Ga_{21.8}$ [50] and $Ni_{50}Mn_{27}Ga_{22}Fe_1$ [25]. This SEM-BSE contrast is interpreted here as a qualitative signature of twin-related refinement.

The contrast in Figure 6b is not uniform, which we attribute to a statistical distribution of nanotwin sizes, as also suggested in Ref. [63]. This distribution may originate from local chemical variations and defects, which can influence the modulation period and twin-formation energetics. Furthermore, due to the finite spatial resolution of the microscope, the observed contrast and apparent band thickness do not necessarily reflect the true twin-size distribution; for instance, a laminate of very thin twins can appear as a single broad grey band, as simulated in Ref. [63]. The microscopy observations thus serve only as a qualitative indication of twin refinement and cannot provide an exact determination of twin size at low temperatures; direct local verification of individual nanotwin boundaries would require TEM or related high-resolution methods.

Nevertheless, the observed refinement is too pronounced to be ascribed solely to changes in, e.g., the lattice parameters or elastic constants of martensite, or to the interfacial energy of the twins—parameters that are typically decisive for the microstructure scaling [71]. Here the average contrast width decreases by nearly an order of magnitude, while none of the above parameters is expected to change dramatically in such a narrow temperature interval, well below the martensite transition temperature.

As these SEM observations are only qualitative, we further estimated the average domain size using XRD and a condition relating twin domain size to diffraction peak merging [72,73]. This condition is satisfied when $m < 2/(sh)$, where $m$ is the twin width in atomic planes, $s$ is the twinning shear, and $h$ is the reciprocal-space coordinate of the measured (h00) reflection. For a typical Ni–Mn–Ga alloy, this critical domain size is around 20 nm [63]. At this threshold, the (400) and (040) reflections merge into a single (400)' line, indicating an apparent shift in diffraction symmetry from monoclinic to pseudo-orthorhombic. This provides an indirect diffraction-based estimate of the average *a/b*-nanotwin size: for measured $a \neq b$, the twins are larger than ≈20 nm, while for measured $a = b$, they are smaller than ≈20 nm.

Figure 7a shows the experimental lattice parameters *a* and *b* in alloy 1 as a function of temperature, alongside the corresponding nanodomain (*a/b*-nanotwin) size versus temperature. The latter relationship was derived from Equation (3) using the experimental $q(T)$ dependence (Figure 2c). The nanodomain size decreases with cooling. The 20 nm domain size is indicated, below which the diffraction peak merging condition is met.

The temperature at which the predicted domain size of 20 nm is reached (using experimental $q(T)$ in Equation (3)) aligns closely with the temperature where diffraction peak merging occurs in XRD experiments (where *a* equals *b* upon cooling). Specifically, these temperatures were about 290 K and 295 K, respectively, a difference that is small within the experimental uncertainties. In other words, the experimental observation of $a = b$ agrees with the ≈20 nm domain size predicted from the experimental $q(T)$ using Equation (3).

In summary, the BSE observations provide a qualitative indication of *a/b*-twin refinement, while the XRD analysis provides indirect diffraction evidence for nanodomains (*a/b*-nanotwins < 20 nm) forming below approximately 290 K and resulting in a diffraction-symmetry shift from monoclinic to pseudo-orthorhombic (within the diffraction resolution). For clarity and consistency with previous works, the pseudo-orthorhombic symmetry is hereafter referred to simply as *orthorhombic*. These observations of twin refinement, *a/b*-nanotwinning, and symmetry shift set the stage for the emergence of LP-C states, which we discuss in the next section.

### *3.3 Long-period commensurate (LP-C) states*

As shown above, nanodomains develop within the evolving incommensurate structure and can be interpreted as emerging *a/b*-nanotwins. As $q$ increases above 2/5, LP-C states can occur for rational $q$, i.e., when an integer multiple (*N)* of the modulation period $p$ aligns with the lattice periodicity, so that $N{\cdot}p$ is an integer number of lattice planes. This commensurability condition—also illustrated in Figure 7b—can be expressed as:

$$p = 5 - 1/N, \text{ where } N = 1,2,3,.., \tag{4}$$

and the length of the long period (*LP*) is:

$$LP = N{\cdot}p = 5N - 1 \tag{5}$$

The determined LP-C states are listed in Table 1, illustrating the relationship between $N$, the modulation period $p$, the modulation vector component $q$, and the corresponding domain size for each state. At one extreme, $N = 1$ yields the theoretically predicted 4O structure [47] ($q$ = 1/2), which has not been experimentally observed in Ni–Mn–Ga. At the other extreme, $N = \infty$ corresponds to $q$ = 2/5, which represents the 10M commensurate structure.

For LP-C states, $q$ and $p$ are rational numbers. These can be written in fractional form using $p = 2/q$ and Equation (4) as:

$$q = \frac{2N}{(5N-1)}, \qquad p = \frac{5N-1}{N}, \tag{6a, 6b}$$

as also listed for $q$ in Table 1.

The domain size of LP-C states is calculated from Equation (3) using the lattice spacing $d_{220}$ = 0.21 nm. The structure naming convention in Table 1 follows Ramsdell notation—commonly used in polytype structures—based on the unit cell size and symmetry. For example, 24O represents an orthorhombic unit cell with 24 stacking planes. An alternative notation is also proposed based on the modulation period $p$ and symmetry, which might be more practical in some cases. Under this scheme, 24O is denoted 4.800O.

The structures with DS > 20 nm (i.e., > 99 lattice planes, N > 20, $q \leq 0.404$) are labeled M (monoclinic) in Table 1. This labeling reflects the diffraction observation that the symmetry remains monoclinic until the diffraction peak-merging condition is reached, at which point it shifts to (pseudo-)orthorhombic (Figure 7c). Consequently, three structural regimes are identified:

(1) Commensurate monoclinic structure for $q = 2/5$ (10M-C).

(2) Incommensurate monoclinic structure for $2/5 < q < 0.404$ (10M-IC).

(3) Orthorhombic structure for $0.404 \leq q \leq 3/7$ (IC and LP-C states including 14O, 24O, ... ),

which broadly correspond to experimental observations (Appendix A1).

The LP-C structures exhibit orthorhombic symmetry. The cases most relevant for this study—14O ($q = 3/7$), 24O ($q = 5/12$), and 34O ($q = 7/17$)—are marked in Figure 7c, and their unit cells are illustrated in Figure 8. The stacking sequence interpretations in the same figure show $2|\bar{2}$ nanotwin boundaries and $(2\bar{3})_2 \leftrightarrow (3\bar{2})_2$ inversions. This reveals that, in addition to aligning with the lattice periodicity, the unit cells in LP-C states also correspond to *a/b*-nanotwins. The specific stacking sequences are: i) $(2\bar{3}2|\bar{2}3\bar{2})$ for 14O, ii) $(2\bar{3}2\bar{3}2|\bar{2}3\bar{2}3\bar{2})$ for 24O, and iii) $(2\bar{3}2\bar{3}2\bar{3}2|\bar{2}3\bar{2}3\bar{2}3\bar{2})$ for 34O, where "|" again represents the *a/b*-nanotwin boundary.

Both studied alloys converge close to $q = 5/12 \approx 0.417$ at low temperatures (Figure 2c, d), which matches the 24O LP-C state (Table 1). The convergence signifies a lock-in transition, in which the modulation becomes fixed at a commensurate value, a phenomenon well-documented in materials exhibiting long periodicity [28,29].

Thus, we identify specific LP-C states, manifesting orthorhombic unit cells that can be viewed as *a/b*-nanotwins. As temperature decreases, the data show convergence of $q$ toward 5/12, consistent with a lock-in to 24O within experimental uncertainty.

### *3.4 Stress response of a/b-nanotwins (q evolution)*

With an understanding of the structural landscape of *a/b*-nanotwins, notably their size and the occurrence of LP-C states, we can further characterize them via the loading response of $q$, i.e., via changes in the *a/b*-nanotwin size. Here, we demonstrate only the initial probing of the stress response; a broader study exceeds the scope of this report.

Following the above reasoning on nanodomain size and Eq. (3), we can link $q$ to *a/b*-nanotwin size and vice versa. In an ideal, perfectly commensurate 10M structure with $q = 2/5$, which lacks inherent inversion of $(2\bar{3})_2$ stacking, the *a/b*-twins are of formally infinite size. In the experiment, *a/b*-twins of size ≈10 µm or smaller are observed [67] (see also Figure 6). Such twin size corresponds to a minuscule deviation of $q$ from 2/5. Saren et al. [70] observed stress-induced redistribution of *a/b*-twins in 10M martensite of $Ni_{50.0}Mn_{28.4}Ga_{21.6}$ under very low applied stress (≈0.1 MPa). The location in the phase diagram (Figure 2a) indicates that this alloy was in a commensurate or near-commensurate state at room temperature. The observed twin sizes ranged from 1 to 10 µm, which, using Eq. (3), corresponds to $2/5 < q < 0.40001$ (i.e., $q$ only marginally above 2/5).

The stress required for twin reorientation increases, and lattice softness decreases when the structure approaches the 14M martensite [25,74]. At that stage, $q >$ 2/5, and the *a/b*-nanotwins refine to a size of a few tens of atomic planes. To examine how such an *a/b*-nanotwinned structure responds to loading, we use the synchrotron tensile-loading data reported in Ref. [75]; the strain was determined from direct $d_{100}$-spacing measurements in a synchrotron transmission diffraction experiment, while the corresponding stress was calculated using the elastic moduli.

The studied sample initially had $q = 0.414$, which is located between the 24O and 34O structures (Figure 7c) and corresponds to an ≈6 nm *a/b*-nanotwin size. Figure 7d illustrates how the structure evolved under increasing strain. Notably, $q$ remained nearly constant until an elastic strain of ≈ 0.15% (≈36 MPa stress), where a transition to a mixed IC-C state occurred. This state consisted solely of the initial incommensurate $q = 0.414$ phase and the commensurate $q = 2/5$ phase. Further strain led to a fully commensurate state, consistent with progressive annihilation of *a/b*-nanotwin boundaries (reduction of boundary density) and the formation of large regions with the long *a*-axis aligned with the applied stress. This threshold-like IC→C response is consistent with the in situ XRD study by Vinogradova et al. [76], who found that IC 10M Ni–Mn–Ga–Fe remained stable under tensile deformation up to a critical strain, above which a mixed IC + C state appeared and the C fraction increased with further deformation.

The specific evolution of $q$ upon loading highlights a qualitative difference between coarse *a/b*-twins and *a/b*-nanotwins. In the near-commensurate regime ($q \approx$ 2/5), micrometre-scale *a/b*-twins redistribute under very low stresses [70]. Such reorientation mainly changes the relative volume fractions of already existing *a*- and *b*-oriented domains without changing the modulation vector $q$. By contrast, in the incommensurate regime with $q \approx 0.414$, where *a/b*-nanodomains are only a few nanometres thick, the domain size is directly linked to $q$ through Eq. (3), so their reorganization requires a change in $q$. However, $q$ is not freely adjustable, because it is also tied to the electronic/lattice-dynamical conditions associated with stabilization of the modulated state [15,16,49,59]. In other words, gradual *a/b*-twin reorientation is possible only when it does not require a change of $q$.

The threshold-like response shown in Fig. 7d suggests that a barrier associated with the coupled modulation/nanotwin configuration must be overcome before local annihilation of nanotwins and the transition into a fully commensurate state can occur. At a still higher elastic strain of ≈0.43% (≈103 MPa), the structure transforms from 10M (q = 2/5) to 14M (q = 2/7) [75]. This intermartensitic transformation can be viewed as a further collective rearrangement of the *a/b*-nanotwin substructure, consistent with a correspondingly higher barrier.

Thus, in the cases considered here, the rearrangement of *a/b*-nanotwins—either within the IC structure or during an intermartensitic transformation—requires stresses two to three orders of magnitude higher than those sufficient for coarse *a/b*-twin-boundary motion. The observed collective response of *a/b*-nanotwins is consistent with the wave–nanotwin duality further developed below.

### *3.5 DFT+U energy landscape of LP-C structures*

Our diffraction experiments and accompanying analysis reveal the existence of the LP-C state 24O, toward which the structure converges upon cooling. To assess the energetic competitiveness of LP-C states across $q$, we use ab initio DFT+U calculations. Given the known limitations of first-principles calculations for Ni–Mn–Ga [77], we first validate the calculations against the reference anharmonic modulation function, Eq. (2).

Figure 9 provides an overview of the DFT+U results. Figure 9a shows the initial validation; it compares the modulation function from Equation (2) with the relative displacements in the 24O structure determined by DFT+U. Except for a slight amplitude difference—dependent on the localization correction parameter U—the displacement profiles are nearly identical. This indicates good agreement between the *ab initio* calculation and our anharmonic modulation description. In contrast, a comparison with a purely harmonic wave in the same plot reveals significant discrepancies. This shows that describing the modulation as purely harmonic or near-harmonic is insufficient, in agreement with earlier *ab initio* calculations. In particular, Gruner et al. [64] demonstrated that harmonic modulation is applicable only for small modulation amplitudes. This is most relevant near the onset of the martensite transformation when the harmonic soft-phonon mode plays a significant role. The modulation is anharmonic for large amplitudes, consistent with relaxation toward NM-like tetragonal distortions within the formed martensite.

Figure 9b shows that increasing the localization correction parameter U results in a systematic decrease in the modulation amplitude. We define amplitude Δ as half the difference between the calculated cell's maximum positive and minimum negative displacement. It relates to the experimental modulation amplitude $A_1$ (Eq. (2)) as $\Delta \approx 1.24 \cdot A_1$. For U = 0, Δ = 11.5% of $d_{220}$, and decreases about linearly with U to Δ = 8.2% of $d_{220}$ for U = 1.8 eV. This range of Δ corresponds reasonably to that observed experimentally, $1.24 \cdot A_1$ = 12.4% for alloy 1 at 10 K and $1.24 \cdot A_1$ = 9.4% for alloy 2 at 2 K. The possible reasons for experimental amplitude variation were discussed in Section 3.1.

To summarize the initial validation, for the considered range of U, both the modulation waveform from DFT+U calculations (Figure 9a) and its amplitude (Figure 9b) agree well with the experimental data and the reference anharmonic modulation function using Eq. (2).

The energy landscape—i.e., a plot of energies for individual commensurate structures with different $q$ values under various Hubbard corrections U—is shown in Figure 9c. To minimize errors due to cell-size effects, the energies are given relative to the energy of non-modulated martensite with the same calculation cell size. The 4O ($p$ = 4) and commensurate 10M ($p$ = 5) structures have been studied before (e.g., Refs. [47,64]). In contrast, the LP-C structures such as 18O ($p$ = 4.500), 14O ($p$ = 4.666), 24O ($p$ = 4.800), and 34O ($p$ = 4.857) have not previously been investigated via ab initio methods.

Two kinds of commensurate $q$ = 2/5 structures were considered in calculations, 10O and 10M. The 10O structure is orthorhombic, corresponding to Eq. (2). Orthorhombic structures have been experimentally observed very close to the martensite transformation and have been previously labeled as 10M' [14,63] (see also Figure 2a). The conventional 10M commensurate structure has a five-layer modulation but additionally exhibits monoclinic distortion. It is typically observed at room temperature, near the martensite transformation [22]. Monoclinic distortion is not directly included in our anharmonic modulation description, as defined by Equation (2). To capture both the anharmonic modulation and the monoclinic distortion of the 10M structure, one can treat these two phenomena independently [78].

A related recent elastic-energy model by Sedlák et al. [54] discusses how the interaction between a modulation wave and an energy landscape favouring NM-like local distortions can produce spontaneous monoclinicity in commensurate 10M. In our DFT+U calculations, the monoclinic distortion can be viewed as an additional relaxation of the 10O structure toward 10M. In the uncorrected (U = 0) calculations, this relaxation lowers the total energy by about 2 meV but produces a monoclinic angle much larger than observed experimentally. Notably, for U = 0.5 eV and higher, the structures and total energies of 10M and 10O martensites become nearly identical; the monoclinicity of 10M is then very small and comparable to experimental observations of $\gamma \approx 90.3°$.

Having described the structures in Figure 9c, we now analyze how their energy profiles vary with U to determine their energetic competitiveness. For U in the range 0–1 eV, the 4O structure exhibits the lowest energy, although this structure has not been observed experimentally in Ni–Mn–Ga. Nonetheless, its predicted low energy is critical for nanotwinning energetics, as discussed above (see also Refs. [63,64] for *a/b*-nanotwin energy calculation).

The value of U = 0.5 eV is sufficient to lower the relative energies of all structures with $p < 4.8$ ($q > 0.4167$) below that of the non-modulated martensite. With further increasing U, additional structures at larger $p$ also fall below the non-modulated martensite. At U = 1.3 eV, there is a shallow but definite minimum in the energy profile for the 14O structure at p = 4.667 ($q = 3/7$). For U = 1.5 eV and U = 1.8 eV, the minimum shifts to the 34O structure at $p = 4.857$. Thus, varying the Hubbard U parameter shifts the energy minimum, indicating that LP-C states are energetically competitive with the commensurate five-layer (10M) structure and other structures such as NM.

The determined trends are primarily qualitative due to the uncertain “right” localization correction parameter U, the narrow energy differences involved, and the use of $Ni_{50}Mn_{25}Ga_{25}$ rather than $Ni_{50}Mn_{28}Ga_{22}$. Moreover, for $p > 4.5$ and U > 1 eV, the energy differences are on the edge of the method accuracy (≈0.1 meV/atom) and even for U = 0 eV, the computed energies remain within a narrow 3.5 meV range. Within these limitations, applying a localization correction brings the LP-C and 10M commensurate states close to degeneracy, with at most a very shallow preference for one LP-C state (14O, 24O, or 34O) depending on U. This sensitivity of the energy gradient to U further supports the sensitivity of the modulation energetics to electronic/magnetic treatment by DFT [46,48,79].

Overall, our DFT+U calculations compare the 0 K total energies of commensurate LP-C structures and show that several of them lie close to the five-layer commensurate (10M) structure, supporting a shallow martensitic energy landscape. The calculated structures also show a systematic increase in tetragonal distortion with increasing $q$ for all considered U values, consistent with a close link between $q$ and tetragonality. However, the finite-temperature selection of specific $q$ values must also involve additional contributions, as discussed in Section 3.7.

### *3.6 Composition–temperature trends in q and LP-C lock-in states*

In this section, we compile the experimental and literature data on the modulation vector component $q$ in the most relevant $Ni_{50}Mn_{25+x}Ga_{25-x}$ compositions and place them in the context of the newly identified LP-C states. Then, we briefly discuss the causes of the modulation evolution.

Considering composition, the alloys near the off-stoichiometric $Ni_{50}Mn_{28}Ga_{22}$, like the two studied here, are of particular practical interest because they exhibit the modulated martensite structure and resulting functionality at room temperature. Meanwhile, the stoichiometric $Ni_{50}Mn_{25}Ga_{25}$ ($Ni_2MnGa$), cubic at room temperature and transforming to martensite below ≈200 K, is highly relevant as the archetypal Heusler alloy.

A graphical overview of the distribution of LP-C states along $q$ is provided in Figure 10a. The compilation of our results and published data on $q$ as a function of temperature is displayed in Figure 10b. These data exhibit composition- and temperature-dependent trends, with $q$ converging to specific values (noting the non-monotonic dependence of $q$ on $x$): 3/7 ($x \approx 0$), 5/12 ($x \approx 3$), and 7/17 ($x \approx 2$). These values correspond to the LP-C states 14O, 24O, and 34O, respectively. The considerable scatter in the data likely reflects the wide range of sources and experimental conditions employed, such as different measurement techniques, sample preparation methods, structure order, and the potential presence of twinning. It is also evident that single-temperature measurements are insufficient since only broader temperature ranges reveal convergence toward specific $q$ values. Taken together, the compiled data indicate composition-dependent low-temperature convergence toward specific rational $q$ values, consistent with LP-C lock-ins.

The alloys studied here, i.e., x ≈ 3, undergo a martensite transformation slightly above room temperature. Near the transformation, at room temperature, the structure is typically commensurate or nearly commensurate ($q \approx 2/5$). Upon further cooling, the structure evolves toward 24O ($q = 5/12$). Measurements by Çakir et al. [23] on an alloy with $x \approx 2$ show $q$ evolving toward the 34O structure ($q = 7/17$).

The particular LP-C state occurring at low temperatures apparently depends on composition. However, the noted non-monotonic dependence of low-temperature $q$ on $x$ suggests that it is also influenced by other factors, such as chemical ordering, residual internal stress, or sample form (powder/single crystal). Table 1 implies that additional LP-C states (e.g., 38O, 58O) are in principle available within the observed $q$ range, and the existing data suggest that some of these may be realized for other compositions. For instance, the trends in the data suggest that $x \approx 1$ could favor a low-temperature state close to 38O.

For stoichiometric alloys $Ni_{50}Mn_{25}Ga_{25}$ ($x \approx 0$), the measured $q$ values cluster just below 3/7, implying that this system prefers to transform into the 14O state. Except for one observation by Singh et al. [59], we are not aware of reports showing $q > 3/7$ for five-layer-derived martensite. In a Ni–Mn–Ga–Fe alloy, $q = 3/7$ (14O) was also observed as the limiting value before transformation to 14M martensite [25].

These composition–temperature trends raise the question of how the observed $q$ evolution and LP-C lock-ins relate to the electronic, lattice-dynamical, and nanotwinning descriptions of Ni–Mn–Ga martensite.

### *3.7 Physical picture of q evolution and wave–nanotwin duality*

In austenite, the tendency toward modulation is commonly associated with Fermi-surface nesting and electron–phonon coupling, often discussed as a charge-density-wave-like (CDW-like) electronic–lattice instability linked to softening of the $TA_2$ phonon branch along [ξξ0] [15,16,81,82]. However, recent electronic-structure calculations indicate that the calculated generalized susceptibility, and thus the predicted modulation wave vector, depend sensitively on the electronic-structure treatment. Therefore, a direct assignment of a specific martensitic $q$ value to a unique Fermi-surface nesting vector should be made with caution [79]. A similar sensitivity appears in the present DFT+U results, where changing U shifts the relative stability of LP-C states with different $q$ values.

Moreover, the softening occurs near $\xi \approx 0.33$, corresponding to the premartensitic 3M/6M modulation rather than directly to the five-layer (5M/10M) modulation of the fully developed martensite. Bungaro et al. [83] showed that, upon tetragonalization of the unit cell toward $c/a \approx 0.94$, ξ shifts continuously toward 0.43.

Combined with the experimental observation of Chulist et al. [14], where a non-modulated $c/a \approx 0.94$ phase appears first and is followed only later by a modulated phase, this suggests that the martensitic modulation may emerge from condensation of the soft $TA_2$ branch in an already tetragonally distorted phase, or during the transition toward it. Although the $\xi \approx 0.33$ instability is commonly associated with so-called *premartensite* [15,83,84], its connection to the actual *martensitic* modulation with $\xi \approx 0.43$ is thus not straightforward. This distinction is especially relevant for Mn-rich alloys such as the present $Ni_{50}Mn_{28}Ga_{22}$ compositions, in which premartensite is not observed at all.

Our DFT+U calculations show that the martensitic energy landscape is shallow and highly competitive. At 0 K, several LP-C structures are energetically close to the five-layer commensurate state. At finite temperature, the selected $q$ must reflect a balance of several energetic contributions, including coupled electronic and lattice-dynamical effects, elastic energy, magnetic interactions, vibrational/entropic terms, and the tendency of the lattice to relax toward NM-like tetragonal distortions. In this view, the coupled electronic and lattice-dynamical effects may bias a favourable modulation periodicity, reflected in $q$, whereas the available low-energy commensurate/LP-C configurations help select the realized $q$ value. This may explain why the martensitic transformation often produces a commensurate or LP-C state rather than a broadly incommensurate modulation.

Mn excess provides one way to connect the composition-dependent soft-phonon anomaly in austenite with the modulation selected in martensite. In austenite, Mn excess shifts the $TA_2$ softening minimum from $\xi \approx 0.33$ in $Ni_2MnGa$ toward lower values, around $\xi \approx 0.30$, and broadens the anomaly due to compositional disorder [16,85]. If a similar composition-induced shift persists along the tetragonalization path into martensite, the resulting *q* in martensite may also tend to decrease with increasing Mn content. This is qualitatively consistent with the transformation-selected *q* values discussed here: $q \approx 3/7$ is observed in $Ni_{50}Mn_{25}Ga_{25}$ after transformation, whereas $q \approx 2/5$ is selected in Mn-rich $Ni_{50}Mn_{28}Ga_{22}$. The broader data summarized in Fig. 10 show, however, that the selected low-temperature LP-C state is not a monotonic function of Mn content. This indicates that *e/a* alone is insufficient and that other factors, such as local chemical or magnetic ordering, may further influence the balance.

Upon further cooling within martensite, the balance between the competing energetic factors can shift. In this respect, the results of Bungaro et al. [83] are particularly relevant: they showed that *q* is closely linked to tetragonality, expressed by the *c/a* ratio. Thus, the increase of *q* upon cooling may generally be related to experimentally observed increasing tetragonal distortion [86]. Other energetic contributions also influence the evolution, but their precise quantitative separation is beyond the scope of the present work. Nonetheless, when another LP-C state becomes energetically favoured, the structure may gradually evolve toward it and eventually lock in.

This preferred-LP-C picture is consistent with the continuum-level approach of Benešová et al. [80], in which an initially five-layer displacement field becomes incommensurate when a seven-layer periodicity emerges as an additional energy minimizer. In the present structural description, this competition is manifested by the drift of *q* away from 2/5 and by lock-in to LP-C states such as 34O, 24O, and ultimately 14O. The 14O state, with $q = 3/7$, is especially important because it lies at the boundary between the five-layer-derived *q*-evolution pathway and the seven-layer periodicity relevant for 14M martensite [25], making the 14O→14M transformation a natural target for future studies.

## 4 Summary and outlook

Our study establishes a comprehensive framework for the thermal and compositional evolution of structural modulation in Ni–Mn–Ga alloys. We show that the evolving modulation wave is associated with emerging *a/b*-nanotwins and long-period commensurate (LP-C) states, thereby linking two descriptions of modulated martensite—wave modulation and nanotwinning—within a single structural picture. Thus, the long-standing wave–nanotwin dispute can be reframed as a true duality: the wave picture captures the structural modulation as reflected by diffraction satellites and $q$ evolution, whereas the nanotwinning picture captures how this modulation is realized locally in the discrete martensitic lattice.

The key findings can be summarized as follows:

- **Anharmonic incommensurate modulation**: Neutron and X-ray diffraction reveal pronounced high-order modulation satellites (up to the eighth order), demonstrating a strongly anharmonic displacement wave. This anharmonicity is quantitatively reproduced by the reference multiharmonic modulation function, enabling precise analysis of the modulation evolution.
- ***a/b*-nanotwinning and structural symmetry**: The evolving incommensurate modulation is associated with $(2\bar{3})_2$ sequence inversions, giving rise to nanodomains that can be interpreted as emerging *a/b*-nanotwins. This structural evolution manifests as a transition from monoclinic to orthorhombic symmetry when the nanodomain size falls below ≈20 nm ($q \approx 0.404$).
- **Long-period structures, energetics, and lock-in transition**: We enumerate a comprehensive set of LP-C structures with rational $q$ values that are simultaneously *a/b*-nanotwins. These LP-C states provide a unifying link between continuous wave modulation and *a/b*-nanotwinning. Specific LP-C structures, namely 14O, 24O, and 34O, are detailed. DFT+U calculations show that LP-C states are energetically competitive with the five-layer commensurate (10M) structure, while experiments and literature data indicate convergence (“lock-in”) to 14O, 24O, and 34O in $Ni_{50}Mn_{25}Ga_{25}$, $Ni_{50}Mn_{28}Ga_{22}$, and $Ni_{50}Mn_{27}Ga_{23}$, respectively.

- **Modulation evolution**: The martensitic transformation appears to select a commensurate or LP-C state close to the locally favoured modulation periodicity, reflected by $q \approx 2/5$ in Mn-rich compositions and $q \approx 3/7$ in stoichiometric $Ni_2MnGa$. Upon further cooling, $q$ evolves within the martensitic free-energy landscape and the structure tends to lock into LP-C states such as 34O, 24O, or 14O, depending on composition. The 14O state (q = 3/7) represents the final waypoint of the five-layer-derived incommensurate pathway before possible transformation to 14M martensite, motivating targeted studies of the 14O→14M transformation.
- **Wave–nanotwin duality:** The observed anharmonicity of the modulation can be interpreted as a signature of relaxation of the continuous modulation wave toward local NM-like tetragonal distortions and associated nanotwinning motifs. In this view, the electronic–lattice instability defines or biases a coherent modulation wave with a preferred periodicity, reflected in $q$, while the discrete martensitic lattice locally accommodates this wave through anharmonic stacking, *a/b*-nanotwin-like inversions, and rational LP-C lock-ins. The modulated martensite is therefore neither a purely wave-like CDW/frozen-phonon state nor a purely nanotwinned (adaptive) structure, but reflects the coupled action of both tendencies.

Taken together, our structural framework provides a basis to interpret the modulation pathway ($q$, $p$, and LP-C lock-ins) across composition and temperature. It also provides a structural starting point for future work connecting this pathway to elastic anomalies and twin-boundary supermobility in Ni–Mn–Ga.

As an outlook, our structural interpretation suggests possible links to dynamic degrees of freedom in the modulated lattice. In particular, phasons—low-energy collective excitations shifting the phase of the modulation wave—could provide a mechanism for local phase slips and modulation adjustment with minimal atomic shuffling [16,87,88]. A recent mechanical model further suggests that such phasons may relax shear loadings in commensurate or weakly incommensurate 10M martensite, while this relaxation channel is suppressed as the modulation becomes strongly incommensurate [54]. This remains a hypothesis in the context of the present work, which provides no direct dynamical evidence for phasons; testing it would require dedicated inelastic-scattering, phonon-calculation, or time-resolved pump-probe studies.

**Acknowledgment**

The authors acknowledge the Institut Laue-Langevin (ILL) for the beam time allocated [5-41-950, 5-15-626], the funding support from the Czech Science Foundation [24-10334S], Ministry of Education, Youth and Sports of the Czech Republic [LUC25051], and the assistance provided by the Ferroic Multifunctionalities project, supported by the Ministry of Education, Youth, and Sports of the Czech Republic [CZ.02.01.01/00/22_008/0004591], co-funded by the European Union. Computational resources were provided by the Ministry of Education, Youth and Sports of the Czech Republic under the Projects e-INFRA CZ [ID:90254] at the IT4Innovations National Supercomputing Center. We thank L. Klimša for assistance with scanning electron microscopy measurements.

## Appendix

### *A1. Five-layered modulated martensite in Ni–Mn–Ga*

The five-layered martensite exhibits a four-level twinning hierarchy, extending from the macro- to the nanoscale [2]. On larger scales, the dominant twins are *a/c*-twins of Type 1 or Type 2, with {101} or approximately {10 1 10} twinning planes. At the mesoscale, {100} compound twins (modulation twins) appear as internal twins within *a/c*-twins, and these can further subdivide into *a/b*-twins, typically observed at length scales of tens of micrometers or less. The *a/b*-twins are {110} compound twins, arising from the slight but non-negligible difference between the lattice parameters *a* and *b*. In the adaptive martensite concept [12,26], the *a*/*b*-twins are internally twinned by nanoscale tetragonal NM twins (<1 nm), often referred to as adaptive nanotwins, Fig. 1e.

Three primary structural cases of five-layer martensite have been identified based on powder diffraction [21–23,89] and high-resolution single-crystal diffraction studies [57,84]:

i) A *monoclinic* commensurate (C) modulated structure with modulation vector component $q \approx 2/5$ (typically observed in $Ni_{50}Mn_{28}Ga_{22}$ at room temperature),

ii) An *orthorhombic* incommensurate (IC) modulated structure with $q \approx 3/7$ (typically observed in stoichiometric $Ni_{50}Mn_{25}Ga_{25}$ at low temperatures), and

iii) A *monoclinic* or *orthorhombic* IC modulated structure with $2/5 < q < 3/7$ (with *q* varying with temperature and composition).

In simple terms, these three cases span from a commensurate 10M martensite close to the martensitic transformation to low-temperature structures whose modulation approaches a seven-layer period.

The symbol $q$ represents the scalar component of the modulation vector that sets the modulation periodicity. In cubic ($L2_1$-frame) reciprocal-space coordinates, the modulation vector is expressed as $\boldsymbol{q} = q \cdot \boldsymbol{g}_{110}^* = (q,q,0)^*$, where $\boldsymbol{g}_{110}^*$ is the $(1,1,0)^*$ reciprocal-space vector. Using the smallest possible face-centered tetragonal unit cell (often denoted as $L1_0$ [90])—or "*diagonal coordinates*"—the modulation vector is expressed as $\boldsymbol{q}_{\mathbf{d}} = q \cdot \boldsymbol{g}_{\mathbf{001d}}^* = (0,0,q)_{\mathrm{d}}^*$ [21]. No published study reports any deviation of the modulation vector from this direction. Hence, characterizing the modulation solely by the component $q$ is equivalent in both coordinate systems. For more information on the concepts of the crystallography of modulated crystals, the reader is referred to Ref. [91].

The modulation vectors $\boldsymbol{q}$ and $\boldsymbol{q}_d$, and the scalar component $q$, are defined in reciprocal space. In direct space (the direct lattice, see also inset in Fig. 2c), the corresponding modulation period $p$ is given in our arrangement by

$$p = 2/q, \tag{A1}$$

measured in number of lattice planes. Note that both $q$ and $p = 2/q$ are defined *relative* to the (110) lattice planes, and are therefore independent of the absolute spacing $d_{110}$. Moreover, $d_{110}$ changes only weakly with temperature [55].

### *A2. Direct-space modulation cases and the adaptive-nanotwin picture*

We illustrate selected modulation cases in direct space in Figure 1, which also serves as a visual link between the wave-like and nanotwinning descriptions. The reference anharmonic modulation function used throughout the paper is given in Eq. (2). Panel (a) shows a commensurate (C), harmonic, short-period modulation, where the modulation period $p$ (expressed in lattice planes) is an integer number and thus alignment with the lattice occurs immediately after the first modulation period. Panel (b) illustrates an incommensurate (IC), harmonic modulation, for which $p$ is irrational and alignment with the lattice is never achieved. Panel (c) depicts an incommensurate, anharmonic modulation, where the waveform deviates from a simple sinusoid through higher-harmonic components. Panel (d) represents a commensurate, anharmonic, long-period modulation, where $p$ is rational but alignment with the lattice requires several modulation periods; this case is referred to here as a long-period commensurate (LP-C) state. Finally, panel (e) introduces the nanotwinning model of the modulated 10M structure, which provides an alternative but widely used structural description: the modulation can be understood as constructed from nanoscale NM tetragonal twins (<1 nm), often termed "adaptive nanotwins" or NM building blocks [12,26]. In this representation, the structure is equivalent to a $(2\bar{3})_2$ basal-plane stacking sequence, and *a/b*-twins arise naturally as inversions of that stacking order [63,64,14].

### *A3. Long-period commensurate states and lock-in transitions*

A commensurate modulation means that the modulation period is in a rational ratio with the lattice periodicity; equivalently, the modulation vector component $q$ (as well as $p = 2/q$) is a rational number. When lattice–modulation alignment occurs only after multiple modulation periods, the structure is referred to as a *long-period commensurate* (LP-C), Fig. 1d. As $q$ evolves between 2/5 and 3/7, many LP-C states are in principle possible, since any non-empty continuous interval of real numbers contains infinitely many rational numbers. This motivates identifying which LP-C states are realized within the experimentally observed $q$ range.

In many modulated systems, LP-C structures can be energetically favored, appear over finite temperature intervals, and in some cases constitute the ground state. With decreasing temperature, the structure can lock into a (meta)stable LP-C state, a process known as a *lock-in transition*. Upon further cooling, repeated lock-ins into different commensurate states can produce staircase-like steps in the $q(T)$ dependence, a phenomenon commonly referred to as the devil's staircase [87,92]. Such a series of

commensurate–incommensurate–commensurate transitions has been well-documented experimentally in many materials with long periodicity along a unique direction [28,29].

The typical progression of a lock-in transition includes: i) a high-symmetry phase transforming into an incommensurately modulated structure, ii) the formation of commensurate regions separated by discommensurations on the microscale as temperature decreases, and iii) a final lock-in into an LP-C modulated phase. However, in many materials, the link between the microscopic rearrangement and the IC→LP-C lock-in transition remains to be elucidated. The situation in Ni–Mn–Ga, where both modulation and twin microstructure evolve strongly with temperature, provides a particularly suitable case to address this missing link.

## Disclosure

The authors report there are no competing interests to declare.

## CRediT authorship contribution statement

L. Straka: Conceptualization; Methodology; Investigation; Formal analysis; Visualization; Writing – original draft; Writing – review & editing; Supervision. P. Veřtát: Conceptualization; Methodology; Investigation (primary experiments); Formal analysis; Visualization; Writing – original draft; Writing – review & editing.
M. Klicpera: Investigation (primary experiments); Writing – review & editing.
O. Fabelo: Investigation (primary experiments); Writing – review & editing. R.Chulist: Investigation (mechanical loading and synchrotron experiments); Writing – review & editing. M. Vinogradova: Investigation (mechanical loading and synchrotron experiments); Methodology; Formal analysis; Writing – review & editing. A. Sozinov: Investigation (mechanical loading and synchrotron experiments); Methodology; Writing – review & editing. M. Zelený: Formal analysis; Investigation (DFT+U); Visualization;

Writing – original draft; Writing – review & editing. P. Sedlak: Validation; Writing – review & editing (lattice dynamics consulting; phason interpretation). H. Seiner: Validation; Funding acquisition; Writing – review & editing (lattice dynamics consulting; phason interpretation); Project administration. O. Heczko: Writing – original draft; Writing – review & editing; Funding acquisition.

**Data availability statement**

The neutron diffraction data are available from Refs. [33,34]. Other data supporting this study's findings are available from Ref. [93].

Figure 1. Schematic illustration of an approximate five-layer modulation in Ni-Mn-Ga martensite. The modulation wave is shown by red curves, lattice with spacing *d* by vertical gray lines, and the modulation period of approximately five layers is marked as *p*. Five representative modulation cases are shown: **a)** *Commensurate, harmonic, short-period* – *p* is an integer multiple of *d*, and the modulation aligns with the lattice after the first period. **b)** *Incommensurate, harmonic* – *p* is irrational multiple of *d* thus alignment with the lattice never occurs. **c)** *Incommensurate, anharmonic* – the modulation wave deviates from a simple sinusoid. **d)** *Commensurate, anharmonic, long-period* – the modulation is anharmonic, *p* is a rational but non-integer multiple of *d*, and alignment with the lattice requires several modulation periods. **e)** Nanotwinning model of 10M martensite (adaptive concept) — the $(2\bar{3})_2$ stacking forming modulation can be viewed as constructed from NM tetragonal nanotwins (<1 nm scale), while *a*/*b*-twins arise from $(2\bar{3})_2 \leftrightarrow (3\bar{2})_2$ stacking inversions.

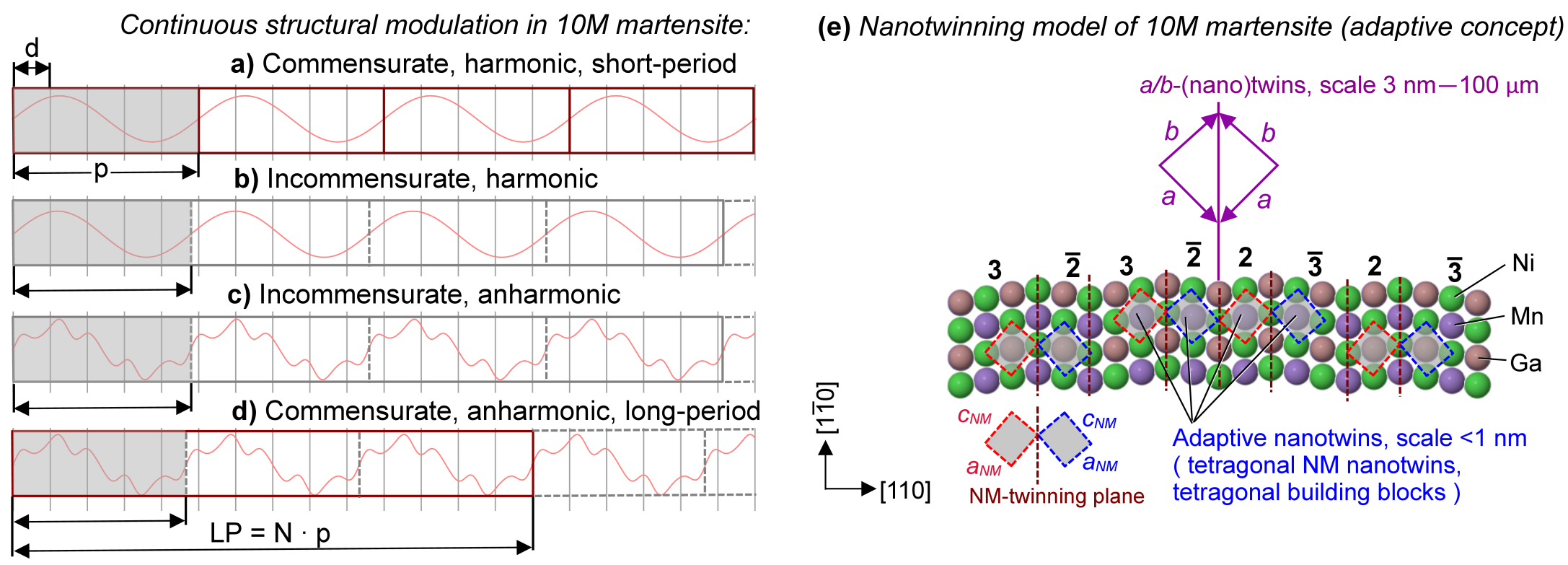

Figure 2. **a)** Schematic phase diagram of the $Ni_{50}Mn_{25+x}Ga_{25-x}$ system, based on Ref. [50], with the measurement paths marked. **b)** Representative neutron diffraction for alloy 2 (300 K → 2 K), showing *q* evolution and the emergence of unusual high-order satellites in the incommensurate structure, marked by red arrows. **c)** The evolution of *q* with temperature for *x* = 2.7, i.e., alloy 1, $Ni_{50.0}Mn_{27.7}Ga_{22.3}$. The inset shows modulation direction and cubic and diagonal coordinates (c ≡ $b_d$ are oriented out of the paper plane). **d)** The evolution of *q* with temperature for *x* = 3.1, i.e., alloy 2, $Ni_{50.0}Mn_{28.1}Ga_{21.9}$. *References to individual regions in (a):* a very narrow region of 10M' near the martensite transformation [14,63], five-layered modulated 10M martensite with *a* > *b* [17–19], nanotwinned martensite with *a* = *b* [50], 14M, NM martensite—seven-layered modulated and tetragonal non-modulated martensites [17,18,27], Saren2020—alloy in which the coarse *a*/*b*-twins were observed optically by Saren et al. [70].

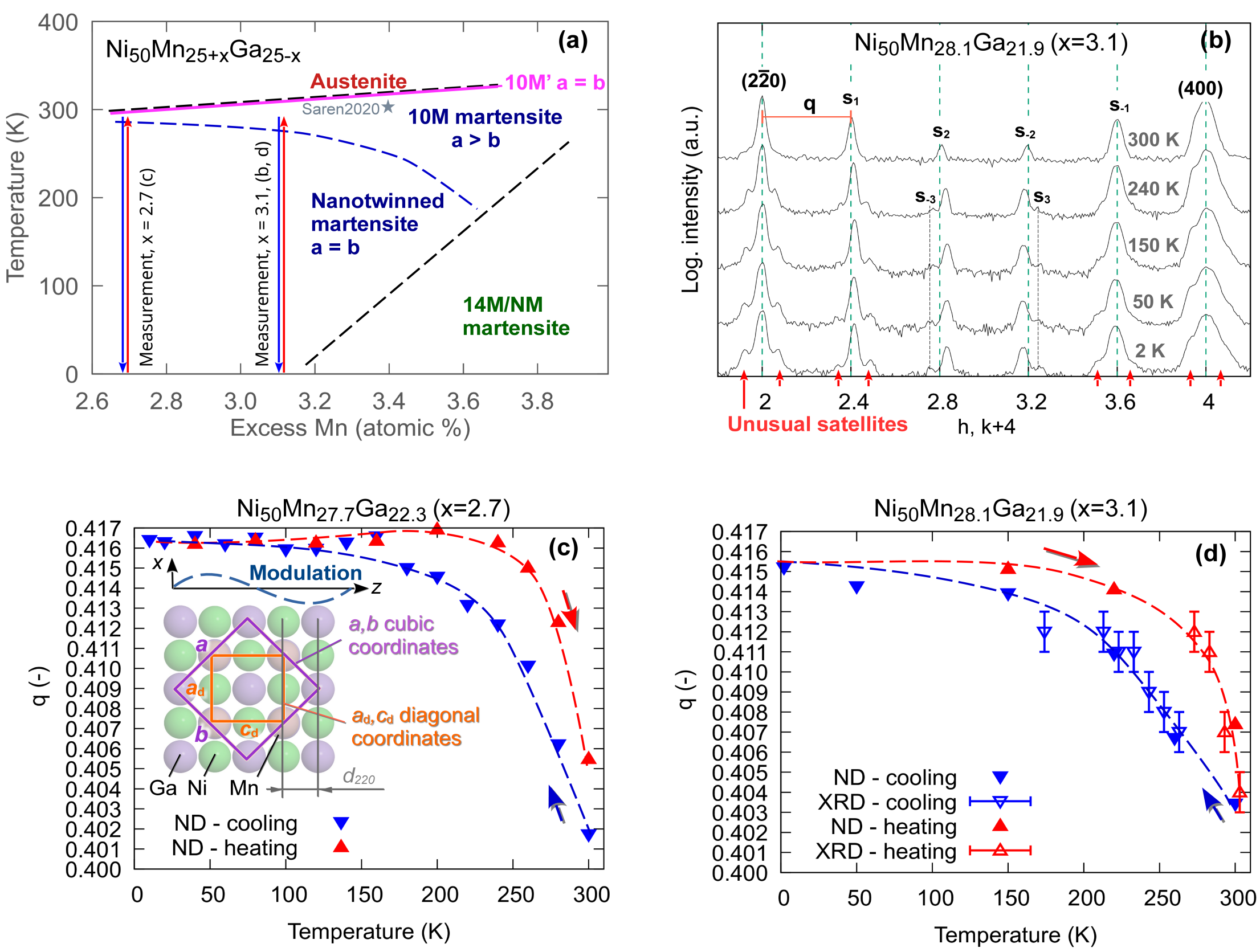

Figure 3. Selected experimental X-ray diffraction patterns showing the evolution of $q$ upon cooling in alloy 2, $Ni_{50.0}Mn_{28.1}Ga_{21.9}$, compared with the model diffraction patterns calculated using the modulation function from Equation (2), at temperatures: **a)** 263 K, **b)** 233 K, **c)** 223 K, **d**) 174 K. High-order satellites from the (220) reflection are marked in (d).

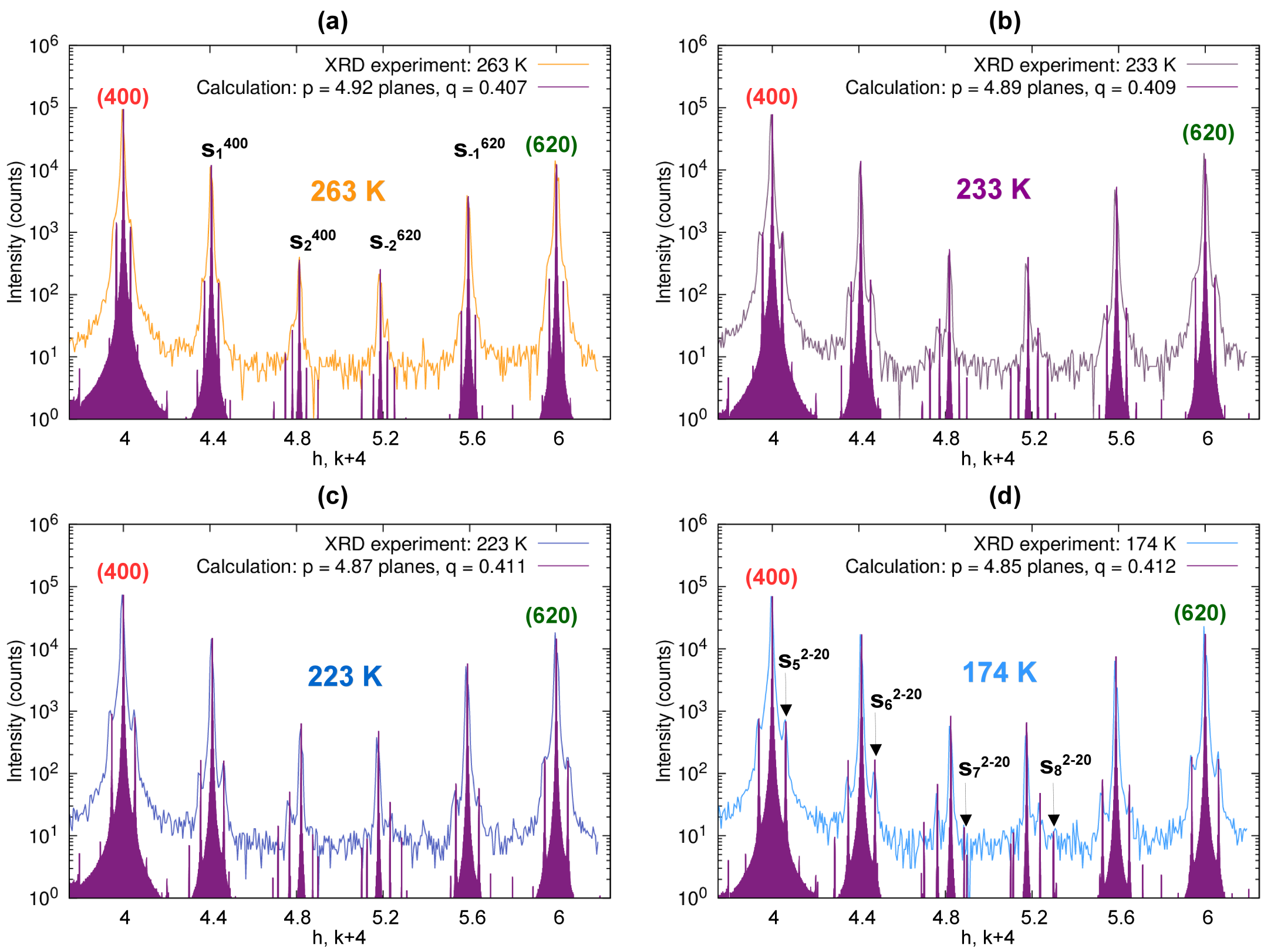

Figure 4. Anharmonic structural modulation with $q = 2/5$, illustrating the interpretation of the displacements as a $(2\overline{3})_2$ stacking sequence: **a)** Modulation function (red curve) juxtaposed with lattice periodicity (vertical lines), determining plane displacements. **b)** Resulting structure, scaled to the lattice. **c)** Resulting structure with displacements threefold exaggerated to highlight the stacking-sequence picture.

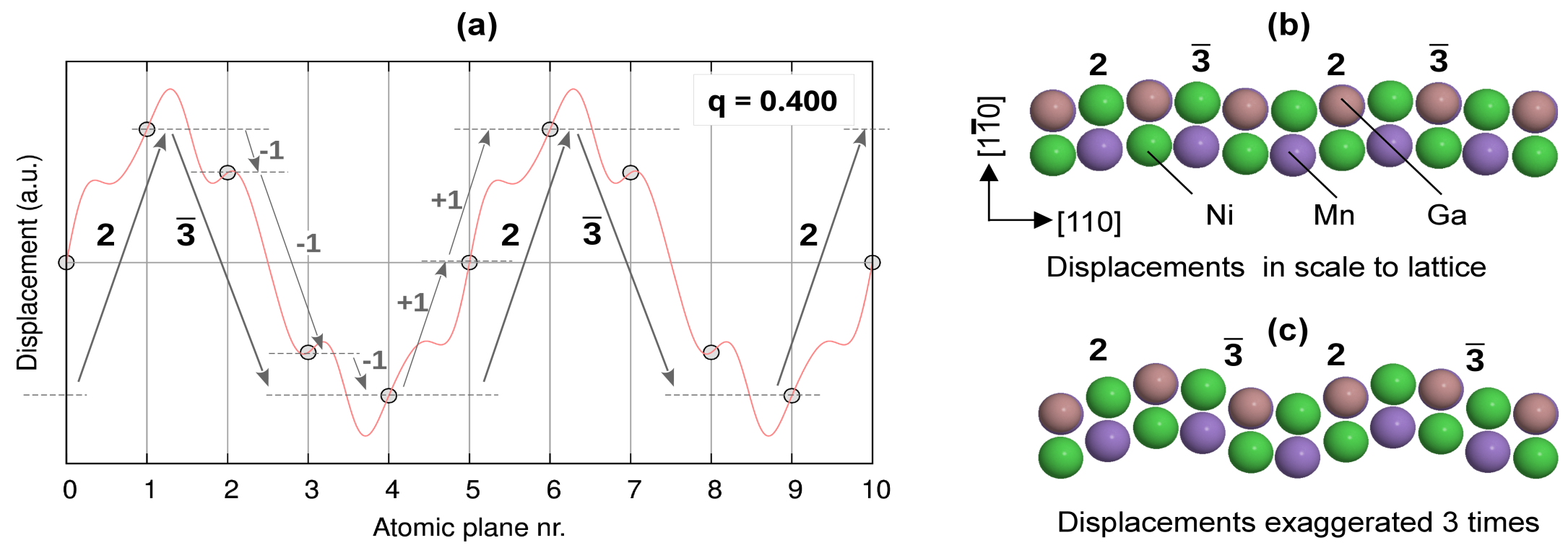

Figure 5. Insights into nanodomain (*a*/*b*-nanotwin) formation arising from the slightly different periodicity of the modulation and lattice. **a)** General principle of wave mixing known, e.g., from acoustics: adding two harmonic signals of different frequency results in a periodic amplitude envelope with beat frequency $|f_1$-$f_2|$. **b)** General principle corresponding to the studied case: interaction of a continuous harmonic wave (modulation function) with a train of narrow pulses (lattice) also results in a periodic envelope. **c)** Application to the studied compound: comparison of commensurate modulation ($q$ = 0.400) with displacements aligned uniformly with the lattice and incommensurate modulation ($q \approx 0.412$), where the mismatch between the modulation and lattice periodicities leads to plane displacements following the envelope. Interpreting the displacements as a stacking sequence reveals reversals between $(2\overline{3})_2$ and $(3\overline{2})_2$ stackings, marking the emergence of *a*/*b*-nanotwin-like nanodomains. The vertical green line marks a $2|\overline{2}$ nanotwin boundary. **d)** A larger-scale depiction for $q$ = 0.400, 0.406, 0.412, revealing different envelope periods, i.e., nanodomain sizes (green dimension lines), and $2|\overline{2}$ nanotwin boundaries (vertical green lines).

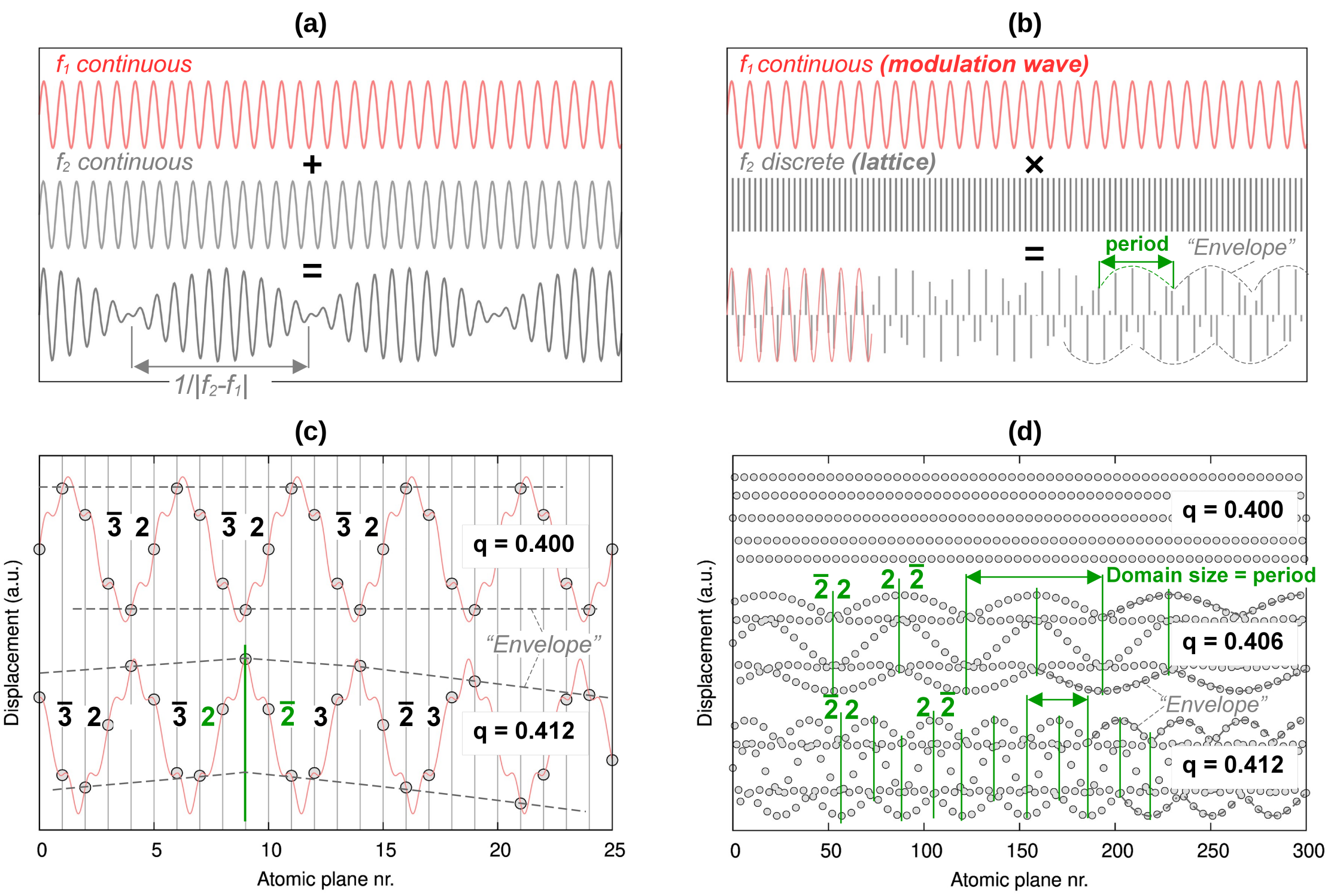

Figure 6. Evolution of the *a/b*-twin-related contrast with cooling as observed in an experiment using scanning electron microscope and BSE contrast in alloy $Ni_{50.0}Mn_{28.2}Ga_{21.8}$ ($x$=3.2). The contrast in the boxed regions has been adjusted post-acquisition to improve visibility. The nearly vertical interface is an *a/c*-twin boundary. **a)** At 298 K, $q \approx 0.400$, broad twin bands oriented horizontally are discernible, some indicated by white arrows. **b)** At 253 K, $q > 0.400$, significantly finer contrast emerges in the same region.

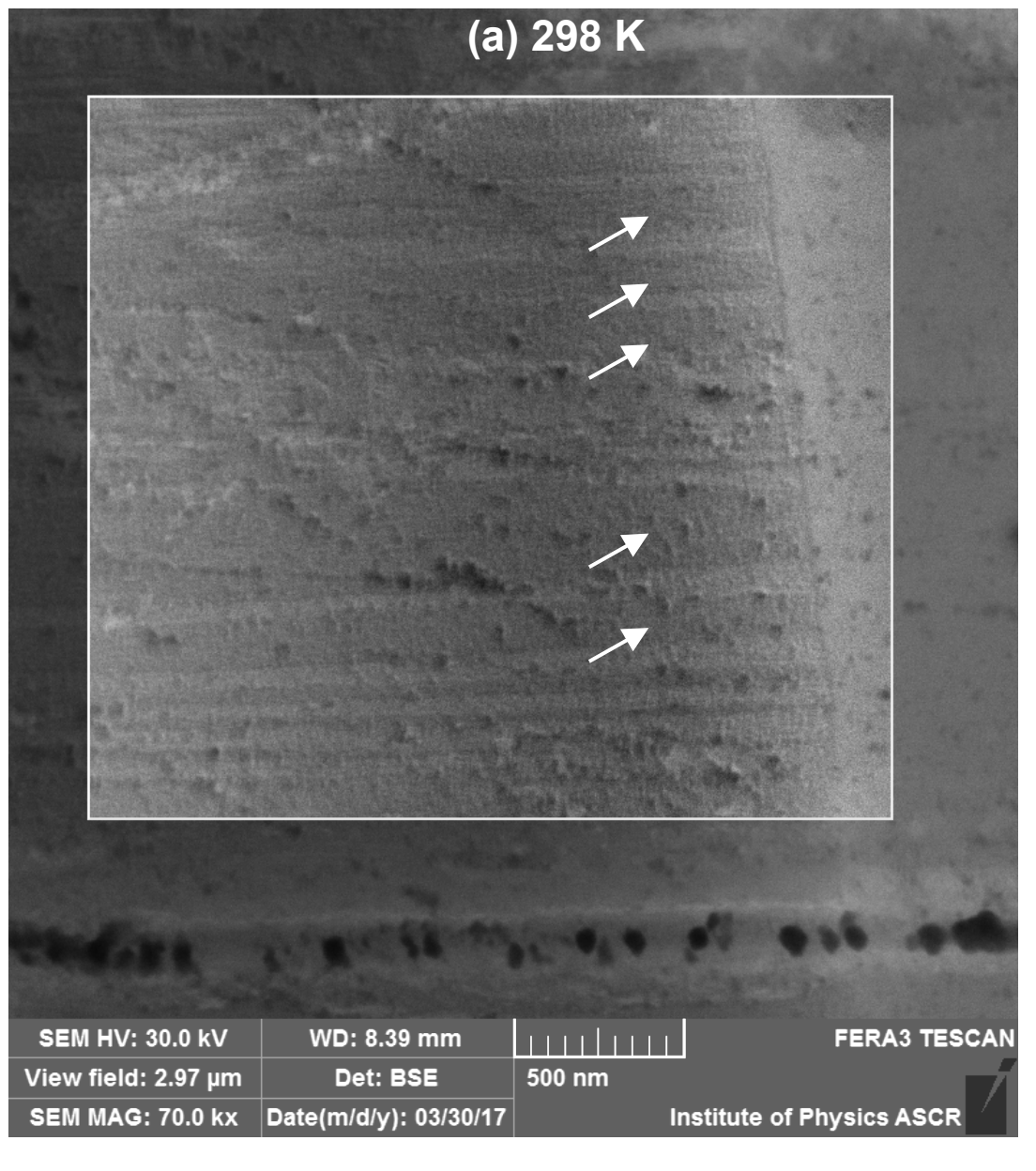

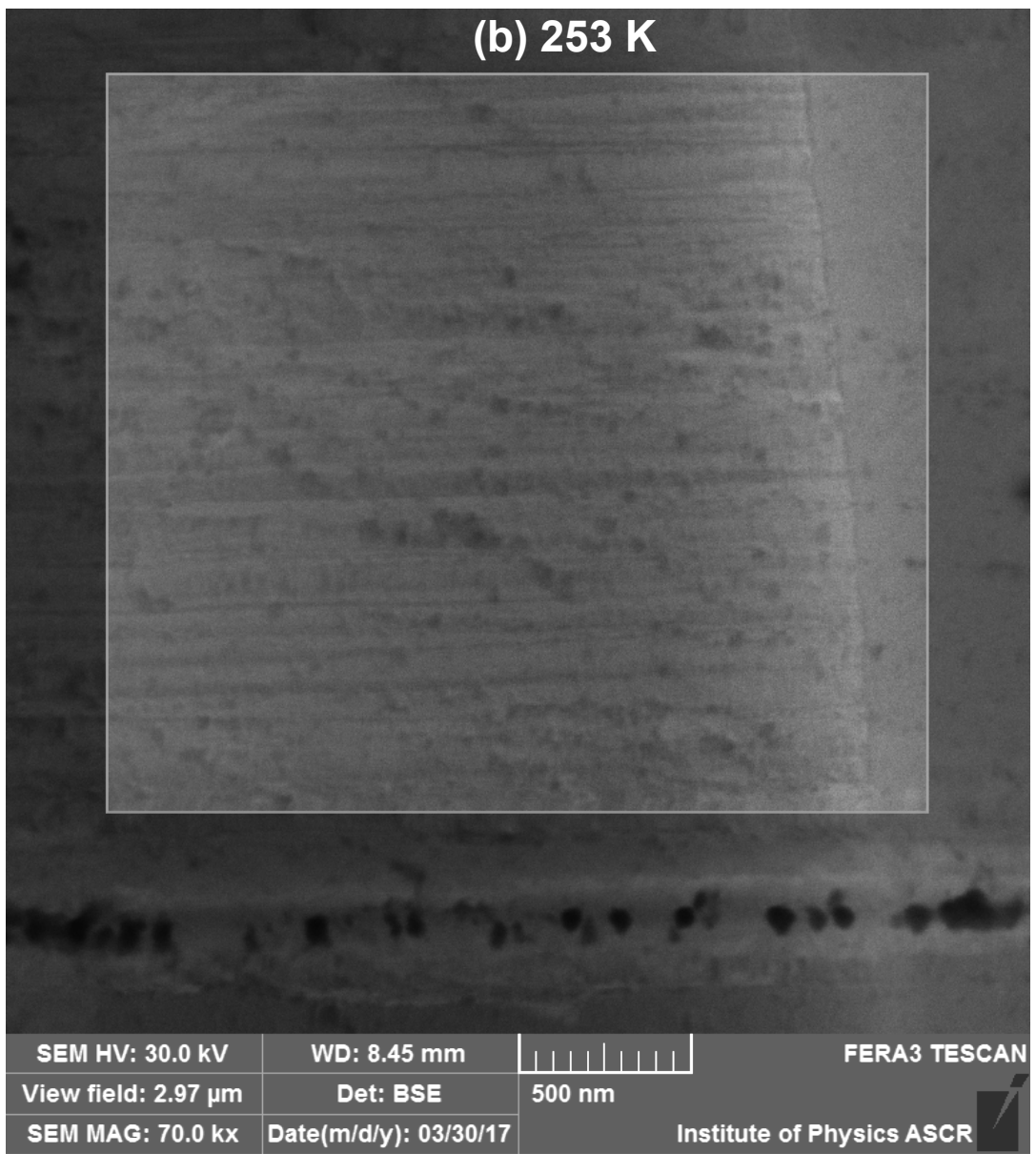

Figure 7. **a)** Relationship between temperature and domain (*a/b*-nanotwin) size, derived from Equation (3) using the measured $q(T)$ dependence in alloy 1, $Ni_{50.0}Mn_{27.7}Ga_{22.3}$. Lattice parameters *a*, *b* (up-triangles and down-triangles, respectively) are charted, and the intervals of monoclinic and orthorhombic symmetry are marked (red and blue colors/regions, respectively) as defined by the diffraction peak merging condition (domain size $DS < \approx 20$ nm, horizontal dashed line). The relation between *a, b* lattice parameters and symmetry is illustrated in the inset. **b)** Representation of long-period state formation using $p = 5 - 1/N$, Equation (4), with $N = 3$ as an example.
**c)** Relationship between domain size and *q* determined using Equation (3) with marked near-room temperature monoclinic structures: 10M commensurate (10M (C)), 10M incommensurate (10M (IC)), and specific low-temperature orthorhombic structures 34O, 24O, and 14O. The dashed magenta line and right axis show the density of $2|\bar{2}$ interfaces (*a/b*-nanotwin boundaries). **d)** Tensile-deformation experiment: *q* is measured at room temperature as a function of tensile strain along the [100] direction (*a*-axis).

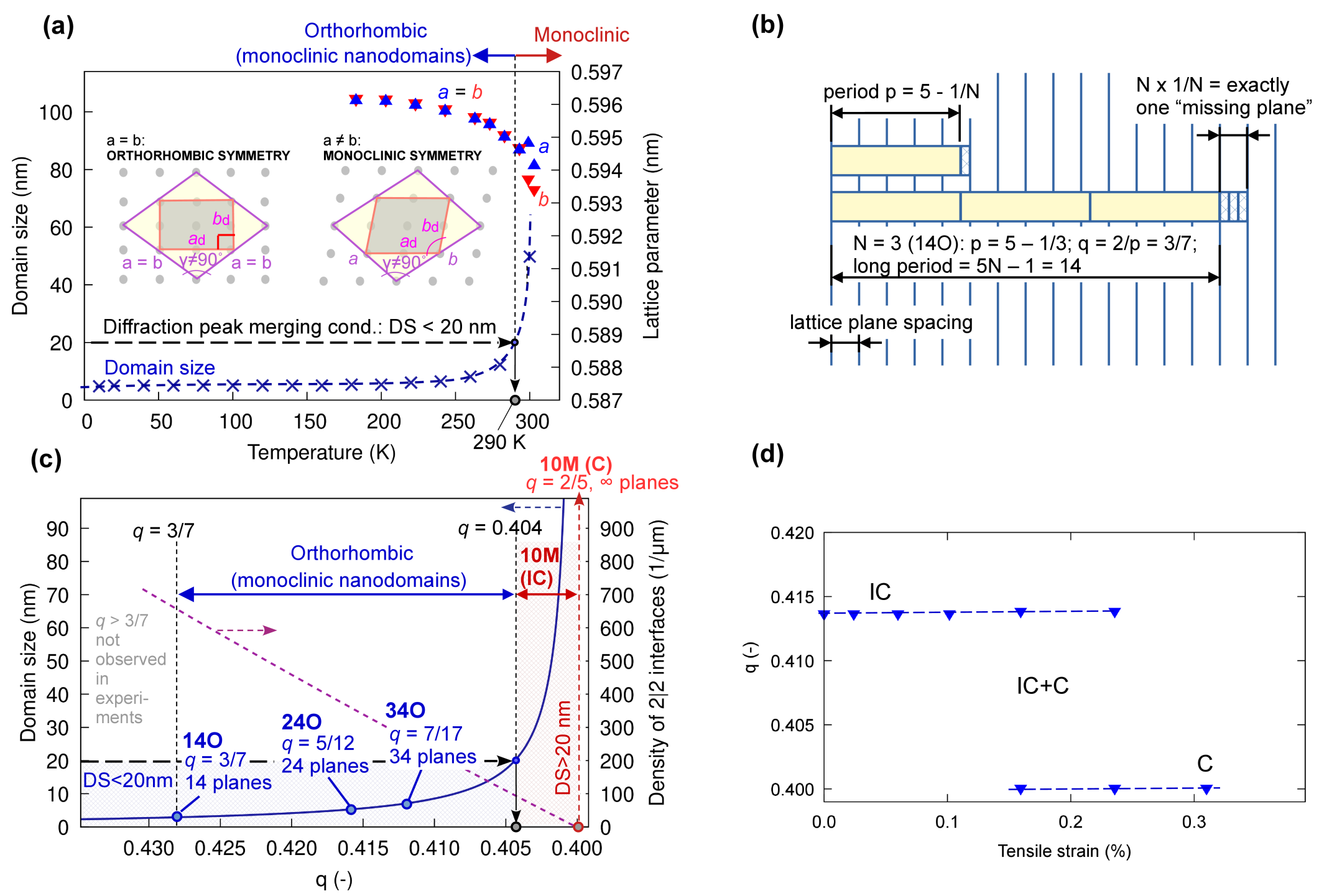

Figure 8. **a)** Unit cell of monoclinic 10M commensurate structure, DFT-calculated. **b, c, d)** Unit cells of orthorhombic structures: 34O (b), 24O (c), and 14O (d), determined using Equation (2), whose unit cells can also be interpreted as *a*/*b*-nanotwins. Modulation displacements were fixed at $A_1 = 10\%$ of $d_{220}$. Horizontal lines aid in visualizing the displacements, and numbers delineating the structures aid in interpreting the displacements as stacking sequences. Vertical green lines mark the twinning planes. Trends with composition and temperature are marked in the top left and bottom left.

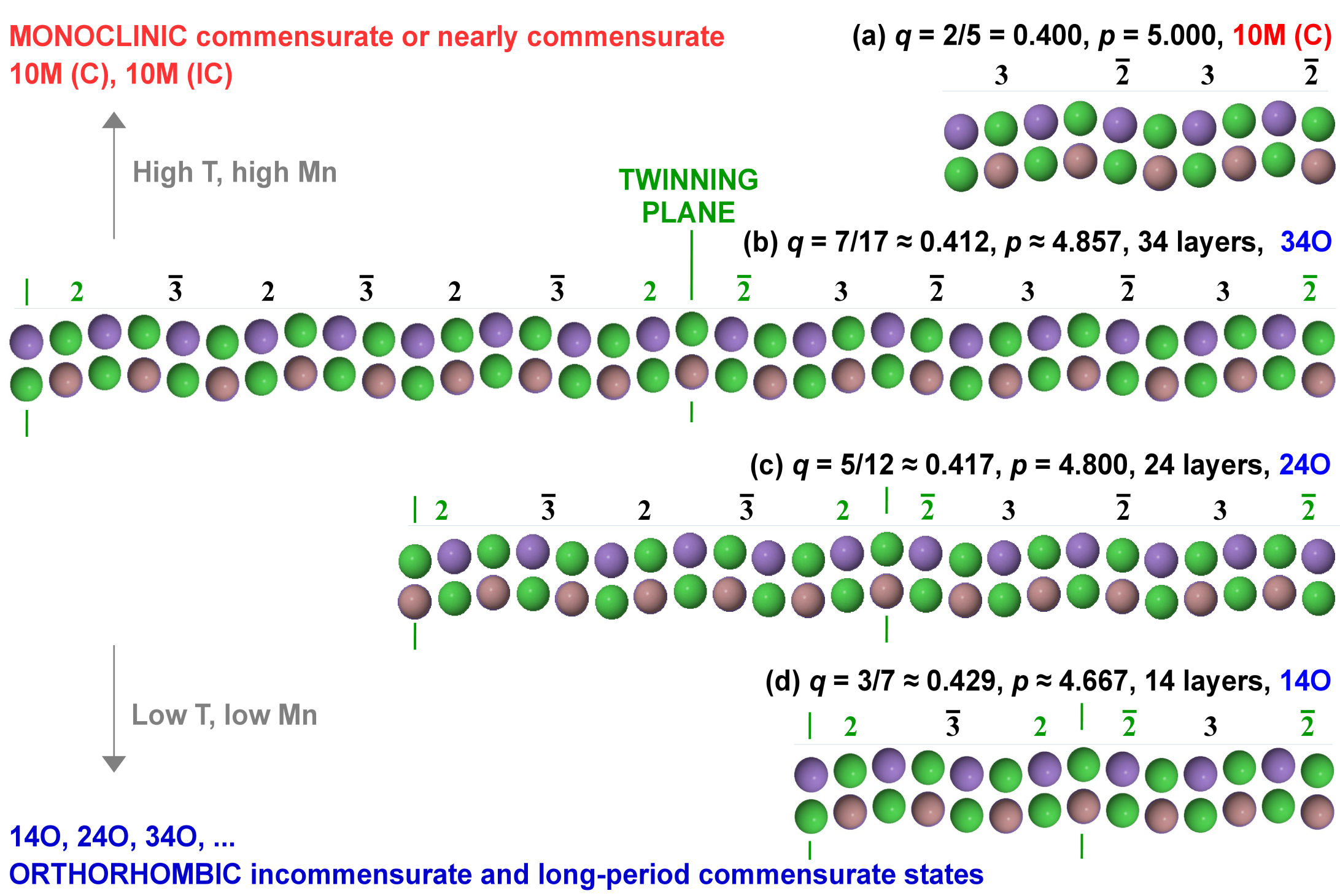

Figure 9. **a)** Comparison between the harmonic modulation (dashed line), the reference anharmonic modulation function (solid red curve according Eq. (2)), and the DFT-calculated displacements (filled symbols) of the (220)-planes in the 24O structure for Hubbard correction parameter U = 0, 0.5, 1.3, and 1.8 eV. Relative displacements exhibit nearly identical waveforms for all U, but maximum displacements (amplitudes) vary slightly, as shown in (b). **b)** Maximum modulation displacements as a function of U and the corresponding experimental maximum modulation displacements determined from ND data for alloy 1 and 2. **c)** Total energy landscape, relative to the non-modulated martensite, as a function of modulation period $p = 2/q$ and U. Individual structures corresponding to a specific modulation period $p$ are marked. The inset illustrates the difference between the 10O and 10M structures: both have similar five-layer modulation but differ in unit cell symmetry (orthorhombic vs monoclinic).

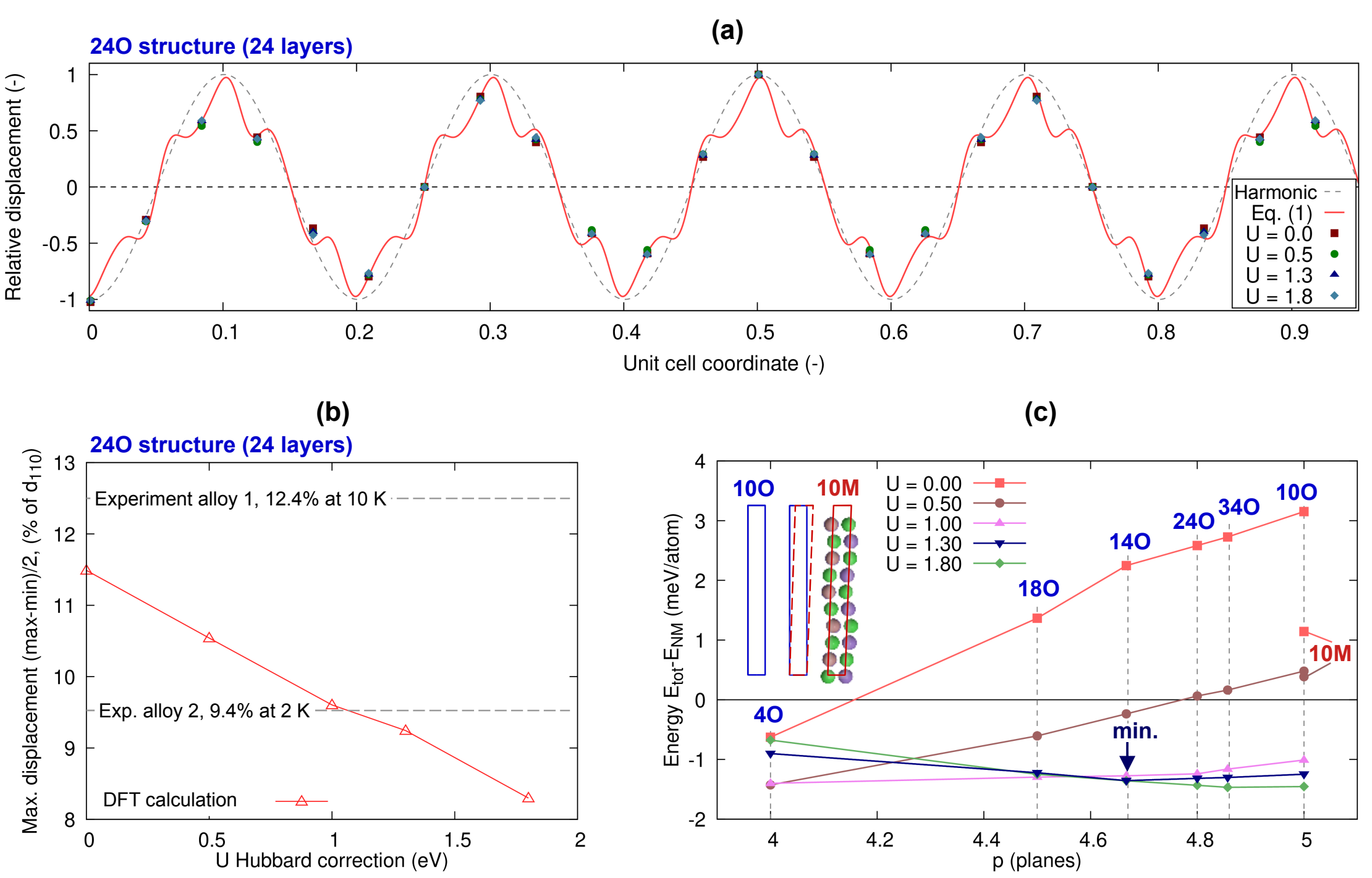

Figure 10. **a)** Distribution of the 10M commensurate, 10M incommensurate, and LP-C states along $q$. **b)** Modulation vector component $q$ as a function of temperature in $Ni_{50}Mn_{25+x}Ga_{25-x}$ alloys. Dashed curves and lines are only guides for the eye. The evolution of $q(T)$, converging toward 3/7, 5/12, and 7/17, suggests low-temperature lock-in to the 14O, 24O, and 34O LP-C states, respectively. Data were compiled from this study and previous literature (see legend): Righi 2007 [22], Fukuda 2009 [35], Kushida 2008 [89], Kushida 2009 [84], Righi 2010 [21], Mariager 2014 [36], Singh 2014 [59], Çakir 2015 [23], and Righi 2021 [24].

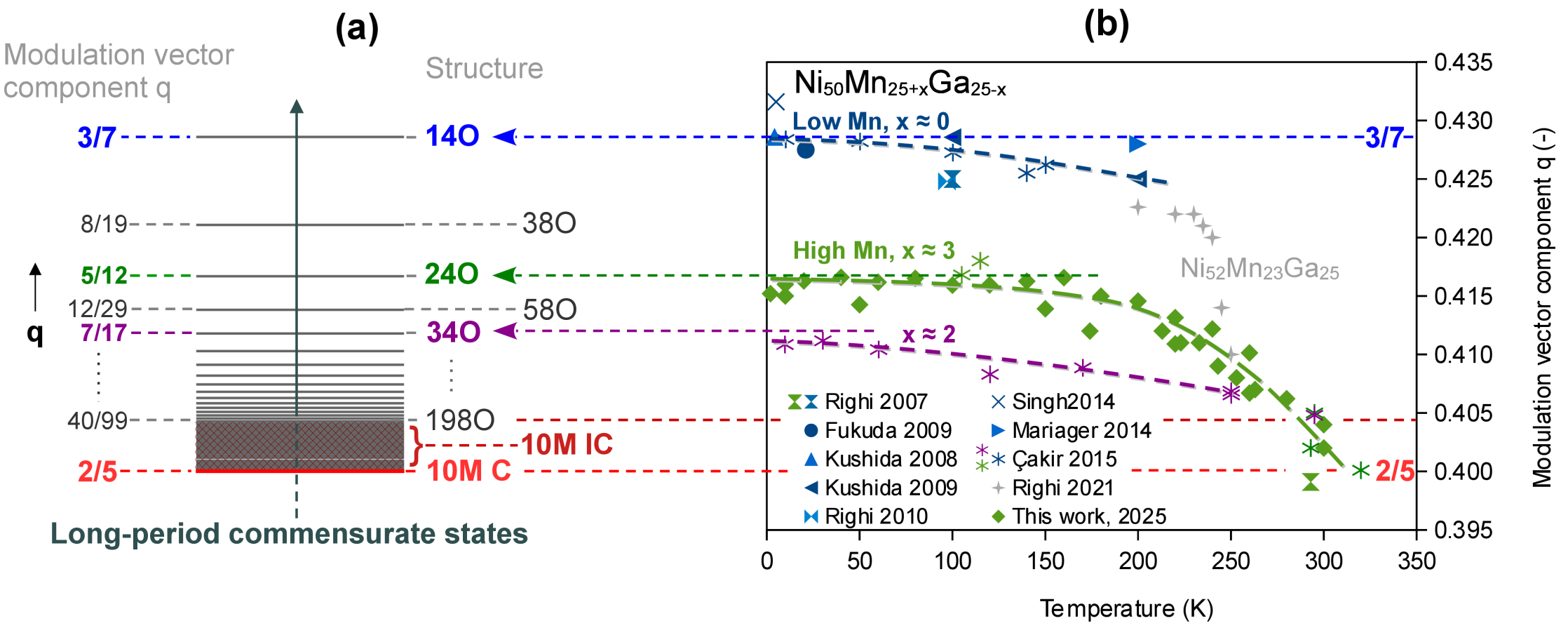

Table 1. LP-C structures derived using Equations (3)–(5).

| N | period p | q = 2/p | q as a ratio | Domain size | | Marking | Alternative | Stacking | Observed in |
|---|---|---|---|---|---|---|---|---|---|
| - | (planes) | (-) | (-) | (planes) | (nm) | | marking | sequence | experiment |
| 1 | 4.000 | 0.5000 | 1/2 (=2/4) | 4 | 0.8 | 4O[*] | 4.000*O* | (22) | no[*] |
| 2 | 4.500 | 0.4444 | 4/9 | 9 | 1.9 | 18O[#] | 4.500*O* | | no |
| **3** | **4.667** | **0.4286** | **3/7 (=6/14)** | **14** | **2.9** | **14O** | 4.667*O* | **(232\|232)** | yes |
| 4 | 4.750 | 0.4211 | 8/19 | 19 | 4.0 | 38O[#] | 4.750*O* | | : |
| **5** | **4.800** | **0.4167** | **5/12 (=10/24)** | **24** | **5.0** | **24O** | 4.800*O* | **(23232\|23232)** | yes |
| 6 | 4.833 | 0.4138 | 12/29 | 29 | 6.1 | 58O[#] | 4.833*O* | | : |
| **7** | **4.857** | **0.4118** | **7/17 (=14/34)** | **34** | **7.1** | **34O** | 4.857*O* | **(2323232\|2323232)** | yes |
| 8 | 4.875 | 0.4103 | 16/39 | 39 | 8.2 | 78O[#] | : | | : |
| 9 | 4.889 | 0.4091 | 9/22 (=18/44) | 44 | 9.2 | 44O | 4.889*O* | | : |
| 10 | 4.900 | 0.4082 | 20/49 | 49 | 10.3 | 98O[#] | 4.900*O* | | : |
| 11 | 4.909 | 0.4074 | 11/27 (=22/54) | 54 | 11.3 | 54O | : | | : |
| 12 | 4.917 | 0.4068 | 24/59 | 59 | 12.4 | 118O[#] | : | | : |
| : | : | : | : | : | : | : | : | | : |
| **> 20** | **> 4.95** | **< 0.404** | **< 40/99** | **> 99** | **> 20** | **10M (IC)**[x] | > 4.950M<br>or 5M (IC) | | yes<br>: |
| : | : | : | : | : | : | : | : | | : |
| **∞** | **5.000** | **0.400** | **2/5** | **∞** | **∞** | **10M (C)**[x] | 5M (C) | **$(23)_2$** | yes |

* Theoretically predicted but not observed in experiment.

# Double cell size to comply with the $L2_1$ order. Ordering ignored for domain size determination.

x Often marked also as 5M when neglecting ordering (C = commensurate, IC = incommensurate); see also discussion on 10O in text.